\documentclass[a4paper,british,12pt]{scrartcl}

\input{prmulticol}

\usepackage{babel}
\usepackage[numbers,sort&compress]{natbib}

\usepackage{amsmath}
\usepackage{amssymb}
\usepackage{color}
\usepackage{graphicx}
\usepackage{xspace}
\usepackage{calc}
\usepackage{array}

\usepackage[small,bf]{caption2}
\usepackage{booktabs}

\usepackage{hyperref}
\usepackage{hypernat}   

\newcommand{\expect}[1]{\ensuremath{\left\langle {#1} \right\rangle}}

\newsavebox{\junk}

\begin{document}

  
\title{The influence of geometry, surface character and flexibility on the
    permeation of ions and water through biological pores}

\author{Oliver Beckstein and Mark S.\ P.\ Sansom%
  \thanks{Department of Biochemistry, University of Oxford, South Parks Road,
    Oxford OX1 3QU, UK. e-mail: oliver@biop.ox.ac.uk,
    WWW: \protect\url{http://sansom.biop.ox.ac.uk/}}} 
\date{12th February 2004}


\maketitle

\begin{abstract}\sf
  A hydrophobic constriction site can act as an efficient barrier to ion and
  water permeation if its diameter is less than the diameter of an ion's first
  hydration shell. This hydrophobic gating mechanism is thought to operate in
  a number of ion channels, e.g.{} the nicotinic receptor, bacterial
  mechanosensitive channels (MscL and MscS) and perhaps in some potassium
  channels (e.g.{} KcsA, MthK, and KvAP). Simplified pore models allow one to
  investigate the primary characteristics of a conduction pathway, namely its
  geometry (shape, pore length, and radius), the chemical character of the
  pore wall surface, and its local flexibility and surface roughness. Our
  extended (ca.~0.1~$\mu$s) molecular dynamic simulations show that a short
  hydrophobic pore is closed to water for radii smaller than 0.45~nm. By
  increasing the polarity of the pore wall (and thus reducing its
  hydrophobicity) the transition radius can be decreased until for hydrophilic
  pores liquid water is stable down to a radius comparable to a water
  molecule's radius. Ions behave similarly but the transition from conducting
  to non-conducting pores is even steeper and occurs at a radius of 0.65~nm
  for hydrophobic pores. The presence of water vapour in a constriction zone
  indicates a barrier for ion permeation. A thermodynamic model can explain
  the behaviour of water in nanopores in terms of the surface tensions, which
  leads to a simple measure of ``hydrophobicity'' in this context.
  Furthermore, increased local flexibility decreases the permeability of polar
  species. An increase in temperature has the same effect, and we hypothesise
  that both effects can be explained by a decrease in the effective
  solvent-surface attraction which in turn leads to an increase in the
  solvent-wall surface free energy.
\end{abstract}


%

\narrowtext

\section{Introduction}
\label{sec:introduction}

Although only about 3~nm thick, the membranes of living cells present an
efficient barrier to polar substances such as water or ions.  Transport of
water and ions into and out of the cell is facilitated by specialised
proteins. For molecules that move down their electro-osmotic gradients the
proteins are ion channels \cite{Hille01} and water pores (aquaporins)
\cite{Fujiyoshi02R}. They share common functional characteristics such as high
specificity (for instance, the potassium channel KcsA prefers potassium ions
over sodium ions, aquaporins allow the flow of water but not of ions or
protons) and transport rates comparable to diffusion in bulk solution. Ion
channels can also be switched (``gated'') between an open (ion conducting) and
closed (ion blocking) state by external signals such as changes in
transmembrane voltage, binding of a ligand, mechanical stress etc.

In recent years a number of near atomic resolution structures of ion channels
\cite{Doy98,Cha98,Zhou01,Bass02,Jiang02a,Jiang03,Kuo03,Miyazawa03} and
aquaporins \cite{Mur00,Fu00,Sui01,Savage03} have been published. Based on
these structures an atomistic understanding of the transport and gating
properties is emerging. If we view transport proteins as molecular machines
which are designed by evolution to perform selective and efficient transport
that can be controlled, i.e. gated, then we can ask what the ``building
blocks'' of these machines are. This involves not only identification and
structural characterisation of the protein domains involved in gating, but
also, at a more abstract level, understanding the underlying physical
principles.

Here we address the questions of what are physical properties that are
important for the permeation of ions and water, and what is their effect? In
particular we will focus on how the flow of ions and water can be controlled,
i.e. we will investigate possible gating mechanisms. This interest is based on
the observation that the putative gate in many known ion channel structures
(in particular, the nicotinic acetylcholine receptor nAChR \cite{Miyazawa03},
the bacterial potassium channels KcsA \cite{Doy98} and KirBac1.1 \cite{Kuo03},
and the mechanosensitive channels MscL \cite{Cha98} and MscS \cite{Bass02}) is
formed by a constriction made from hydrophobic residues. For example, the
closed state nAChR structure displays an ion pathway which is still wide
enough (radius $R\approx0.31$~nm) to admit three water molecules (radius of a
water molecule $r_{w}=0.14$~nm) or one potassium ion
($r_{\mathrm{K}^{+}}=0.133$~nm) with half of its first hydration shell intact.
It is somewhat surprising that a pore does not have to be completely
physically occluded to prevent the flow of ions. On the other hand ions can
readily move through the KcsA \cite{Doy98,Zhou01} selectivity filter although
its radius is less than 0.15~nm. When an ion enters the filter it has to shed
its hydration shell at a high cost in free energy---the solvation free energy
for a potassium ion is about $-320$~kJ\,mol$^{-1}$ \cite{Lyubartsev98}.  The
filter is lined by backbone oxygen atoms which coordinate the potassium ion
and substitute for its hydration shell, thus reducing the desolvation barrier
of $+320$~kJ\,mol$^{-1}$ to about $+12$~kJ\,mol$^{-1}$ \cite{Berneche01}. The
putative gates differ from the selectivity filter in that they are lined by
\emph{hydrophobic} residues. For nAChR \cite{Unw93,Miyazawa03} and MscL
\cite{Cha98,Moe00} it was already hypothesised that these residues cannot
substitute for water molecules so that the energetic cost of desolvation
prevents the passage of the ion.

It appears that the \emph{geometry} of a pore (its radius, length and shape)
and the \emph{chemical character of the pore wall} have a great influence on
the permeation of ions and water. In addition, \emph{local flexibility} (i.e.\ 
fluctuations in the protein structure as opposed to concerted larger scale
motions) of the pore lining might play a role, as seen in simulations of
K$^{+}$ permeation through KcsA \cite{Berneche01}.

It seems difficult to comprehend the dynamical nature of transport phenomena
from static crystal structures alone. Computational methods can be used to
complement the experimentally observed picture. In particular, classical
molecular dynamics (MD) can be used to investigate the behaviour of water or
ionic solutions in the environment presented by a protein.
This realistic environment, however, makes it difficult to disentangle the
contributions of various pore properties. In order to be able to reduce the
number of parameters simplified pore models can be designed to capture the
characteristics in question
\cite{Lyn96,All99,Beckstein01,Crozier01,Allen02,Allen03b,Beckstein03}. We
focus on the influence of geometry, pore wall character and flexibility. In
previous studies we investigated the behaviour of water in these
``nanopores''. In particular, we found that below a critical radius liquid
water becomes unstable in the pore and the pore is predominantly filled with
water vapour \cite{Beckstein03}. We hypothesised that a pore environment which
cannot sustain liquid water would also present a high energetic barrier to an
ion \cite{Beckstein01}. Although it seems plausible that absence of water
would imply the absence of ions it has not been demonstrated previously.

In this work we explicitly consider ions in model pores and explore
dimensions, pore wall character, and flexibility in more detail than in our
previous work. We also present a simple thermodynamic model based on surface
energies that explains the observed behaviour of water in hydrophobic pores.

\section{Methods and Theory}
\label{sec:methods}

\subsection{Model}
\label{sec:model}

\begin{figure*}[bt]
  \centering
  \includegraphics[width=0.5\linewidth,keepaspectratio,clip]{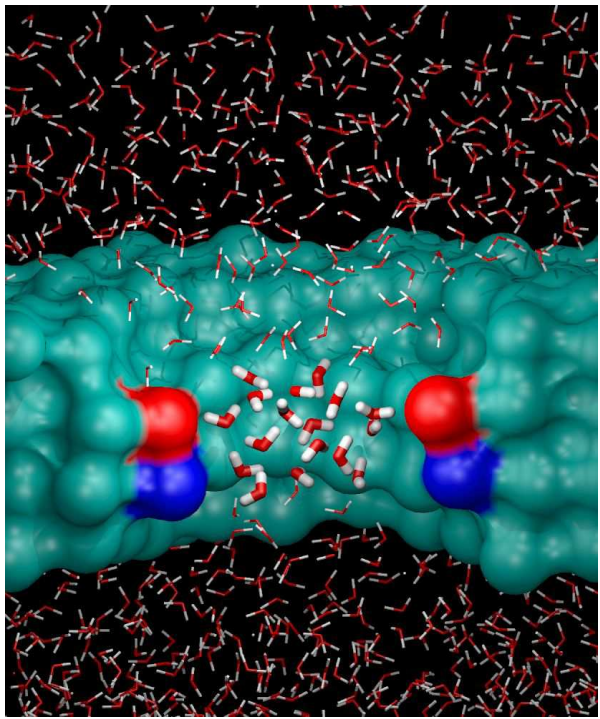}
  \caption{Simulation system: The solvent-accessible
    surface of a model pore with pore radius $R=0.55$~nm is shown in light
    blue (parts of the pore close to the observer are removed for clarity).
    The pore is not completely hydrophobic due to two dipoles (red/blue) in
    the pore wall which have the same magnitude as the peptide bond dipole.
    Water molecules in the pore are depicted in licorice representation and as
    lines in the mouth region or the bulk (image created with VMD \protect\cite{VMD}
    and Raster3D \protect\cite{Raster3D}).}
  \label{fig:model}
\end{figure*}

The pore models were constructed as described previously \cite{Beckstein03}.
Briefly, they consist of concentric rings of methane-like pseudo-atoms of van
der Waals radius $0.195$~nm (figure~\ref{fig:model}). These are held in their
equilibrium position by harmonic constraints with spring constant $k_{0}=1000\ 
\mathrm{kJ}\, \mathrm{mol}^{-1}\, \mathrm{nm}^{-2}$. A pore consists of two
mouth regions (length $0.4$~nm, i.e. one layer of pseudo atoms, and radius
$1.0$~nm) at either end of the constriction site of length $L$ (varied between
$0.4$~nm and $2.0$~nm) and radius $R$ (varied between $0.15$~nm and $1.0$~nm).
The pore was embedded in a membrane mimetic, a slab of pseudo atoms held on a
cubic lattice with unit cell length $0.39$~nm with harmonic springs of
strength $k_{0}$.
Pores with a polar surface were created by placing partial charges of $\pm0.38
e$ on atomic sites $0.2$~nm apart. The resulting dipole moments pointed
parallel to the pore axis with a magnitude of $3.6$~D each, which is
comparable to the dipole moment of the peptide bond (ca.\ $3.7$~D
\cite{Eisenberg79}).
%
%

\subsection{Molecular Dynamics}
\label{sec:meth-md}

MD simulations were performed with \textsc{gromacs} v3.1.4 \cite{Lindahl01}
and the SPC water model \cite{Hermans84}. The Lennard-Jones-parameters for the
interaction between a methane-like pseudo-atom and the water oxygen are
$\epsilon_{\text{CO}}=0.906493$~kJ~mol$^{-1}$ and
$\sigma_{\text{CO}}=0.342692$~nm; parameters for sodium and chloride ions are
taken from the \textsc{Gromacs} force field.
The integration time step was $2$~fs and coordinates were saved every $2$~ps.
With periodic boundary conditions, long range electrostatic interactions were
computed with a particle mesh Ewald method (real space cutoff $1$~nm, grid
spacing $0.15$~nm, 4th order interpolation \citep{Dar93}) while the short
range van der Waals forces were calculated within a radius of $1$~nm. This
treatment of electrostatic effects is known to influence the permeation of
ions, especially in narrow pores. These artifacts need to be accounted for in
order to obtain quantitative predictions of ionic currents \cite{AllenT04}.
However, though most likely present in our case, too, they are less important
because of our emphasis on principles and order-of-magnitude effects.  The
neighbour list (radius $1$~nm) was updated every 10 steps.
Weak coupling algorithms \cite{Ber84} were used to simulate at constant
temperature ($T=300$~K, time constant $0.1$~ps) and pressure ($P=1$~bar,
compressibility $4.5\times10^{-5}$~bar$^{-1}$, time constant $1$~ps) with the
$x$ and $y$ dimensions of the simulation cell held fixed at $4$~nm. The total
thickness of the water reservoir in the $z$-direction was $3.0$~nm, ensuring
bulk-like water behaviour far from the membrane mimetic.
The initial system configuration was created by solvating the slab-embedded
pore model with water. For simulations with ions, some water molecules were
replaced with ions to reach the target concentration. Initially there were
always ions present in pores with $R \geq 0.4$~nm.
A typical simulation box measured $3.9\times 3.9\times 4.6$~nm$^{3}$ and
contained about $1500$ water molecules and $280$ pseudo atoms (and between 25
to 28 Na$^{+}$ and Cl$^{-}$ ions each).

\subsection{State based analysis}

We define discrete states by mapping equilibrium states of the whole system
(snapshots from the MD equilibrium trajectory) onto numbers $\omega_{i}$.  The
same approach is used in statistical mechanics to map different microscopic
states to one macroscopic state, characterised by the value of a state
variable, here called $\omega$.  This approach is used to label the two phase
states that the pore water exhibits in our simulations. We either find
liquid-filled pores or vapour-filled ones. Due to the small pore volumes
``vapour'' typically refers to zero or one water molecule in the cavity. We
use the density in the pore as an indicator of the phase state. As described
previously \cite{Beckstein03} we assign the state using a Schmitt-trigger
procedure \cite{Sakmann83} in order to avoid spurious state changes due to
fluctuations from the interfacial region.  When the water density $n(t)$ rises
above $0.65$ of the density of bulk water, $n_{0}$ ($n_{0}=1.0$~g\,cm$^{-3}$
at $T=300$~K and $P=1$~bar) the liquid state, i.e.{} $\omega=1$, is assigned
to the phase state at time $t$. When $n(t)$ drops below $0.25 n_{0}$ the
vapour state ($\omega=0$) is assigned.
The pore in the liquid-filled state is termed ``open'' because our simulations
show that significant amounts of water pass through it; furthermore, we also
demonstrate in this work that a pore that sustains a liquid environment
potentially allows ions to permeate. A vapour-plugged pore, however, will
prevent ion permeation and is said to be ``closed''.

Liquid-filled pores and vapour-filled pores are assumed to be in equilibrium.
The ``openness'' (or probability for the occurrence of the liquid, i.e.{}
open, state is
\begin{equation}
    \expect{\omega} = \frac{1}{T_{\mathrm{sim}}} \int_{0}^{T_{\mathrm{sim}}}\!\!dt\,
    \omega(t) = \frac{T_{o}}{T_{\mathrm{sim}}}, 
\end{equation}
where $T_{\mathrm{sim}}$ denotes the total simulation time, $T_{o}$ the total
time that the pore is \emph{open}, whereas $T_{c}$ is the total time in the
vapour or closed state.  If $\expect{\omega}>\frac{1}{2}$ then equilibrium is
on the side of liquid (and vapour is a meta stable state), otherwise the
stable phase state is vapour.
The equilibrium is governed by the equilibrium constant
\begin{gather}
  \begin{split}
    K(R) &= \frac{T_{c}(R)}{T_{o}(R)} = \frac{T_{\mathrm{sim}}-T_{o}(R)}{T_{o}(R)}\\
         &= \expect{\omega(R)}^{-1}-1
  \end{split}
\end{gather}
which is trivially related to the openness. %
[Note that for $T_{o}=0$ or $T_{c}=0$ it is meaningless to compute $K$ because
it indicates that the simulation time was too short to sample any state
changes.]

\subsection{Thermodynamic model for liquid-vapour equilibrium in pores}
\label{sec:tdynmodel}

We investigate a subsystem of the whole simulation system comprising of the
pore of volume $V=L \pi R^{2}$. The subsystem can exchange water molecules
with the bulk water outside the pore, which acts as a particle reservoir at
average chemical potential $\mu$. $\mu$ is implicitly determined by the
constant average density of water in the bulk system, which in turn is
prescribed by simulating at constant pressure.  Hence we can use the grand
potential $\Omega(T,V,\mu) = -p V$ as the free energy to describe the
subsystem in the closed and open state \cite{Allen03b}. 
The equilibrium constant $K$ is determined by the free energy difference
between the closed (vapour) and open (liquid) state,  
\begin{gather}
  \label{eq:OpennesstoFE}
  \begin{split}
    \beta\,\Delta\Omega(R) &= \beta\,[\Omega_{c}(R) - \Omega_{o}(R)] 
                           = -\ln K(R) \\
           &= -\ln \bigl(\expect{\omega(R)}^{-1} - 1\bigr),
  \end{split}
\end{gather}
using $\beta=1/k_{B}T$. Following \citet{Evans90}, we write the free energy difference between
the vapour and the liquid state with the corresponding surface contributions
as
\begin{multline}
  \Delta\Omega(R) = \Omega_{v}(R) - \Omega_{l}(R) \\
  = -p_{v}(T,\mu)\, L\,\pi R^{2}  + 2\pi R\, L\: \gamma_{vw} + 2\pi R^{2}
  \gamma_{lv} \\
  -\bigl(-p_{l}(T,\mu) L\, \pi R^{2} + 2\pi R\, L\:  \gamma_{lw}\bigr)
\end{multline}
Here the index $w$ indicates the solid pore wall, $l$ the liquid and $v$ the
vapour phase; for instance $\gamma_{lv}$ is the surface tension or surface
free energy per area of the liquid-vapour interface.  The system is fairly
close to bulk phase coexistence so we can expand the pressure around the
saturation chemical potential $\mu_{\mathrm{sat}}(T)$ in a Taylor series
\begin{multline}
    p(T,\mu) = p(T,\mu_{\mathrm{sat}}) \\
     + (\mu-\mu_{\mathrm{sat}})\:
    \frac{\partial p(T,\mu)}{\partial\mu}\bigg|_{\mu=\mu_{\mathrm{sat}}} +
    \cdots,
\end{multline}
where at saturation $p_{l}(T,\mu_{\mathrm{sat}}) =
p_{v}(T,\mu_{\mathrm{sat}})$.  This leads to a simple parabolic form for the
free energy difference between the two states:
\widetext
\begin{gather}
\begin{split}
  \label{eq:DOmega}
  \Delta\Omega(R) &= \Bigl[2 \gamma_{lv} - \bigl(\mu-\mu_{\mathrm{sat}}\bigr)
  \bigl(n_{v}(T,\mu_{\mathrm{sat}}) - n_{l}(T,\mu_{\mathrm{sat}})\bigr) L
  \Bigr]\pi R^{2}
  + 2\pi L (\gamma_{vw} - \gamma_{lw}) R \\
  &= 2\Bigl[\gamma_{lv} + \frac{1}{2}\Delta\mu\,\Delta
  n_{vl}\,L\Bigr]\pi\:R^{2} + 2\pi L\, \Delta\gamma_{w}\: R,
\end{split}
\end{gather}
\narrowtext
\noindent
where we define the distance of the state from saturation $\Delta\mu := \mu -
\mu_{\mathrm{sat}}$, the difference in densities $\Delta n_{vl} := n_{l} -
n_{v}$ at saturation, and the difference in surface free energies of the two
phases with the wall, $\Delta\gamma_{w} := \gamma_{vw} - \gamma_{lw}$.  The
term $\Delta\mu \Delta n_{vl}\, L$ is small for $L<10$~nm as the system is
close to phase coexistence (for $L\approx 1$~nm it is about $10^{-3}$ times
smaller than $\gamma_{lv} = 17\ k_{B}T \mathrm{nm}^{-2}$ when estimated from
$\Delta n_{vl}\,\Delta\mu\,L \approx \Delta P\,L \approx 1\ \mathrm{bar}\times
L = 2.4\times 10^{-3} k_{B}T \mathrm{nm}^{-2} \times L$, $T=300$~K) and will
be neglected. Only the difference between the surface free energies enters the
model so we express it as the contact angle $\theta_{e}$, using the
macroscopic definition from Young's equation $\gamma_{vw} - \gamma_{lw} =
\gamma_{lv}\cos\theta_{e}$ (see, for instance, Ref.~\citep{Gen85}). Then
equation~\ref{eq:DOmega} becomes
\begin{equation}
  \label{eq:DOmegaTheta}
  \Delta\Omega(R,L,\theta_{e}) = 2\pi\,R\,\gamma_{lv}(R + L\,\cos\theta_{e}).
\end{equation}
(equation~\ref{eq:DOmegaTheta} is similar to the simple model derived by
\citet{Allen03b} but it includes $\gamma_{vw}-\gamma_{lw}$ and hence
$\theta_{e}$ instead of just $\gamma_{lw}$.)  For fixed pore length $L$ and a
given pore material, characterised by $\theta_{e}$, the graph of
$\Delta\Omega(R)$ over the pore radius $R$ describes a parabola containing the
origin.
Although it is not \emph{a priori} obvious that such a macroscopic treatment
(equation~\ref{eq:DOmega} or equation~\ref{eq:DOmegaTheta}) of a nanoscale
system is 
meaningful the MD results for the free energy difference presented in
section~\ref{sec:toymodels-water} can be adequately explained within this
model.

The connection between the model free energy $\Delta\Omega(R)$ and the
behaviour of the system as observed in MD simulations, i.e.{} the openness
$\expect{\omega(R)}$, is established by inverting
equation~\ref{eq:OpennesstoFE}, yielding
\begin{equation}
  \label{eq:FEtoOpenness}
  \expect{\omega(R)}=\frac{1}{1+\exp[-\beta\,\Delta\Omega(R)]}.
\end{equation}
For linear $\Delta\Omega(R)$ this represents a sigmoidal curve but non-linear
terms (as in equation~\ref{eq:DOmega}) change its shape considerably.

\section{Results and Discussion}
\label{sec:results}

First we investigate the behaviour of pure water in nanopores and quantify the
influence of geometry, pore surface and local flexibility on the equilibrium
between liquid and vapour filled pores.  Then simulations for a NaCl
electrolyte are analysed and compared to the pure water case. The intention
behind these studies is to explore more fully the ways in which a hydrophobic
gate may be opened, and how other factors might modulate such hydrophobic
gating. This is particularly important in the context of current models for
the gating of nAChR \cite{Unw00,Miyazawa03} that suggest that the transition
between the closed and open states of the pore involves both an increase in
pore radius and in polarity. We also note that recent discussion of the state
(open vs.\ closed) of MscS have suggested a hydrophobic gating mechanism
\cite{Anishkin04}, similar to that proposed for nAChR \cite{Beckstein01}.

\subsection{Pure water}
\label{sec:toymodels-water}

A large number of simulations were run for water in model pores (total
simulation time $>5\ \mu$s) in order to investigate the influence of pore
radius ($0.15\ \mathrm{nm} \leq R \leq 1\ \mathrm{nm}$) and pore surface
character (hydrophobic vs.\  amphipathic vs.\ hydrophilic). Local
flexibility (wall atom positional root mean square deviations (RMSDs) $0.039\ 
\mathrm{nm}\leq \rho \leq 0.192\ \mathrm{nm}$) and temperature dependence in
the range $273\ \mathrm{K} \leq T \leq 450\ \mathrm{K}$ were examined for a
hydrophobic $R=0.55$~nm pore.

\paragraph{Pore dimensions and surface character}

%
\begin{figure*}[btp]
  \centering
  \includegraphics[width=0.8\linewidth,keepaspectratio,clip]{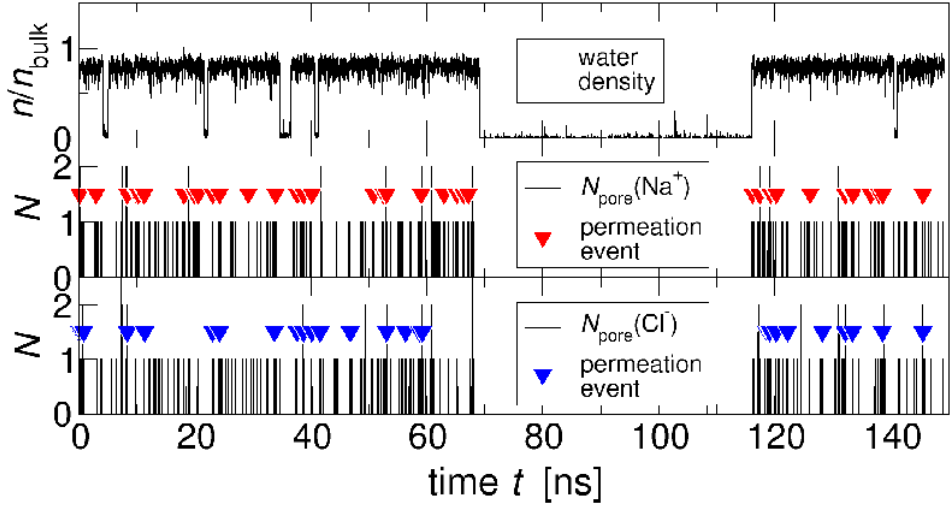}
  \caption{Liquid-vapour oscillations of water in a $R=0.65$~nm apolar pore
    (top panel) when bathed in a $1.3$~M NaCl solution. As indicated by the
    number $N$ of ions within the pore, sodium (middle) and chloride ions
    (bottom) are only observed in the pore when there is also liquid water
    ($n/n_{\mathrm{bulk}} \approx 0.8$) present.  Permeation events are
    indicated by triangles; ions do not permeate the pore during the vapour
    phases.}
  \label{fig:states}
\end{figure*}
\begin{figure*}[btp]
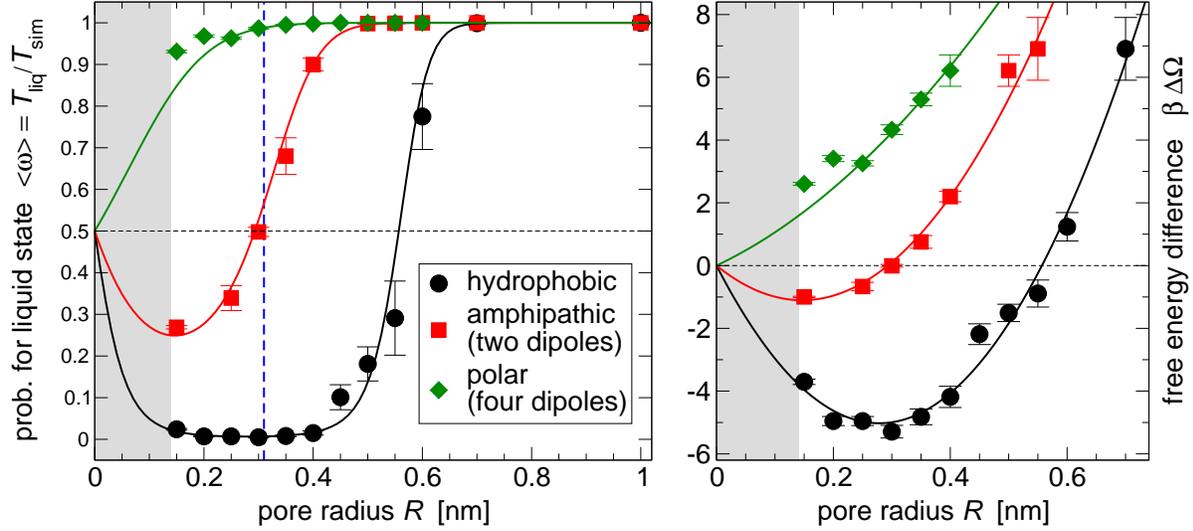

  \centering
  \includegraphics[clip,height=7cm]{figs/omega}\rule{1em}{0mm}%
  \includegraphics[clip,height=7cm]{figs/FE}
  \caption{Water in model pores. Left: openness $\expect{\omega}$ over
    radius. Right: free energy difference $\Delta\Omega$ (in $k_{B}T$) between
    vapour and liquid state. The grey region indicates radii smaller than the
    radius of a water molecule ($0.14$~nm).  Data points are obtained from MD
    simulations with the errors estimated from block averages
    \protect\cite{Hess99}. The continuous lines are fits of the model
    (equation~\protect\ref{eq:DOmega}) to the data points (right) or the openness
    computed from the model (equation~\protect\ref{eq:FEtoOpenness}).  The vertical
    line indicates the radius of the closed nAChR gate ($R=0.31$~nm).}
  \label{fig:omegafree}
\end{figure*}

The water density in short nanopores oscillates between liquid and vapour on a
nanosecond time scale as seen in figure~\ref{fig:states}, a manifestation of
capillary evaporation and condensation at the nanoscale \cite{Beckstein03}. In
figure~\ref{fig:omegafree} the openness and the free energy difference
$\Delta\Omega$ between vapour and liquid states is shown for different pore
surfaces. For hydrophobic and amphipathic pores a strong dependence of the
pore state on the radius is apparent. The stable thermodynamic state switches
from vapour ($\Delta\Omega < 0$ or $\expect{\omega} < \frac{1}{2}$) to liquid
($\Delta\Omega > 0$ or $\expect{\omega} > \frac{1}{2}$) at a critical radius
$R_{c} = -L \cos\theta_{e}$ (using $\Delta\Omega(R_{c})=0$ in
equation~\ref{eq:DOmegaTheta}). $R_{c}$ is $0.56\pm0.03$~nm for the
hydrophobic pore and $0.29\pm0.02$~nm for the amphipathic one. The functional
form equation~\ref{eq:DOmega} fits the data from the MD simulations
(table~\ref{tab:tdynmodel}) well (the continuous lines in
figure~\ref{fig:omegafree}).
%
\begin{table*}[bt]
  \caption{\label{tab:tdynmodel}Parameters of the thermodynamic model
    equation~\protect\ref{eq:DOmega} fitted to the MD results and resulting contact
    angle $\theta_{e}$. Experimental values for
    $\gamma_{lv} = 72\times 10^{-3}\ \mathrm{J} \mathrm{m}^{-2} =
    17\,k_{B}T\,\mbox{nm}^{-2}$ 
    and $\theta_{e} \approx 118^{\circ}$ for water on a flat methyl
    (--CH$_{3}$) terminated self assembled monolayer
    \protect\cite{Wu02,Whitesides90}  
    indicate that the  simple thermodynamic model gives the 
    right order of magnitude results. (All values at room temperature)}
  \centering
  \begin{tabular}{lll>{$}r<{$}@{$\pm$}>{$}l<{$}
                     >{$}r<{$}@{$\pm$}>{$}l<{$}
                     >{$}r<{$}@{$\pm$}>{$}l<{$}}
  \toprule
    \multicolumn{3}{l}{pore surface character} 
    & \multicolumn{2}{c}{$\gamma_{lv} +  
                          \frac{1}{2} \Delta\mu\, \Delta n_{vl}\, L$} 
    & \multicolumn{2}{c}{$\Delta\gamma_{w}=\gamma_{vw} - \gamma_{lw}$} 
    & \multicolumn{2}{c}{$\theta_{e}$} \\
        & &
    & \multicolumn{2}{c}{$[k_{B}T\,\mbox{nm}^{-2}]$} 
    & \multicolumn{2}{c}{$[k_{B}T\,\mbox{nm}^{-2}]$} 
    & \multicolumn{2}{c}{$[^{\circ}]$} \\
  \midrule
    bare CH$_{4}$ & \emph{0} & ``hydrophobic'' & +10   & 1    & -7.2 & 0.2 
                             & 134 & 3 \\
    two dipoles  & \emph{D2} & ``amphipathic'' & +8.2  & 0.5  & -3.0 & 0.2
                             & 111 & 2 \\
    four dipoles & \emph{D4} & ``hydrophilic'' & +2.8  & 0.5  & +1.8 & 0.2
                             &  51 & 10 \\
  \bottomrule
  \end{tabular}
\end{table*}
%
%
%
The coefficient of the quadratic term, $\gamma_{lv} + \frac{1}{2} \Delta\mu\,
\Delta n_{vl}\, L$, is positive and similar for both the hydrophobic and the
amphipathic (two dipoles, abbreviated \emph{D2}) pores
(table~\ref{tab:tdynmodel}), consistent with the model
equation~\ref{eq:DOmega}, which predicts this coefficient to be independent of
the pore wall. 
For the polar pore (four dipoles, \emph{D4}) the data is more ambiguous.  Only
if the first two data points at small radii are excluded from the fit as
outliers is this coefficient positive (but still three times smaller than the
one for the less hydrophilic pores). The outliers show that a high density of
local charges leads to a higher probability of the pore being liquid-filled
than predicted by the macroscopic model, possibly indicating a shortcoming of
the model to subsume inhomogeneous potentials into the surface tension terms.
For our system parameters, the coefficient is in fact dominated by the water
liquid-vapour surface tension $\gamma_{lv}$ (see section~\ref{sec:tdynmodel})
and hence we will use its value as an approximation to $\gamma_{lv}$.
$\Delta\gamma_{w}$, the difference in surface tensions between the wall and
vapour or liquid, becomes more positive with increasing polarity of the pore
wall.  It effectively measures the hydrophobicity of the wall.  This becomes
even more apparent when the (macroscopic) contact angle $\cos\theta_{e} =
\Delta\gamma_{w}/\gamma_{lv}$ is formally computed
(table~\ref{tab:tdynmodel}).  Macroscopically, a hydrophobic surface can be
defined as one with $\theta_{e}>90^{\circ}$. This allows us to call the apolar
pore ($\theta_{e}=134^{\circ}$) ``hydrophobic'' compared to the
``amphipathic'' pore \emph{D2} ($\theta_{e}=111^{\circ}$; still hydrophobic
but with some ``hydrophilic'' patches).  A ``hydrophilic'' pore like the
\emph{D4} system ($\theta_{e}\approx 51^{\circ}$) is characterised by
$\Delta\gamma_{w}>0$ or $\theta_{e}<90^{\circ}$ and liquid is always the
preferred phase in the pore, regardless of $R$.

Experimental macroscopic contact angles $\theta_{e}$ for water on flat methyl
(--CH$_{3}$) terminated self assembled monolayers are reported up to
$118^{\circ}$ \cite{Wu02,Whitesides90}. A ``microscopic contact angle'' of
$135^{\circ}\pm15^{\circ}$ was calculated for a droplet of 90 water molecules
(radius $R\approx 0.75$~nm) on a methyl terminated film from MD simulations
\cite{Hautman91}, consistent with our result of $\theta_{e}=134^{\circ}$. MD
simulations seem to overestimate the experimental equilibrium contact angle.
In this context we note that although the macroscopic equilibrium contact
angle is only established about $10$~nm from the contact line, the local contact
angle $\theta_{l}$ in the core structure of the contact line (in the region
$3$--$10$~nm from the contact line) in water (and electrolytes) is predicted
to be \emph{larger} than $\theta_{e}$ due to electrostatic screening effects
\cite{Gen85}.

From the model free energy the openness can be calculated
(equation~\ref{eq:FEtoOpenness}) and displayed together with the MD results
(figure~\ref{fig:omegafree}). Because the model effectively treats the liquid
as a structureless continuum it is not meaningful to extrapolate to radii
smaller than the radius of a water molecule (the shaded region in
figure~\ref{fig:omegafree}). Nevertheless, the MD results lie on the model
curve even in the direct vicinity of this region. Further tests of
equation~\ref{eq:DOmegaTheta} by varying the length of the pore together with
the radius confirm the model qualitatively (data not shown).

Our model implies that for both nano- and mesoscale pores the cost of creating
the liquid-vapour interface is the only force driving the filling of a
hydrophobic ($\Delta\gamma_{w}\leq0$) pore. Little free energy $\Delta\mu
\Delta n_{vl}\, L\pi R^{2}$ is gained by creating a bulk-like liquid in the
pore instead of vapour.

\paragraph{Local flexibility}

\begin{figure*}[tb]
  \centering \includegraphics[clip,width=0.8\linewidth]{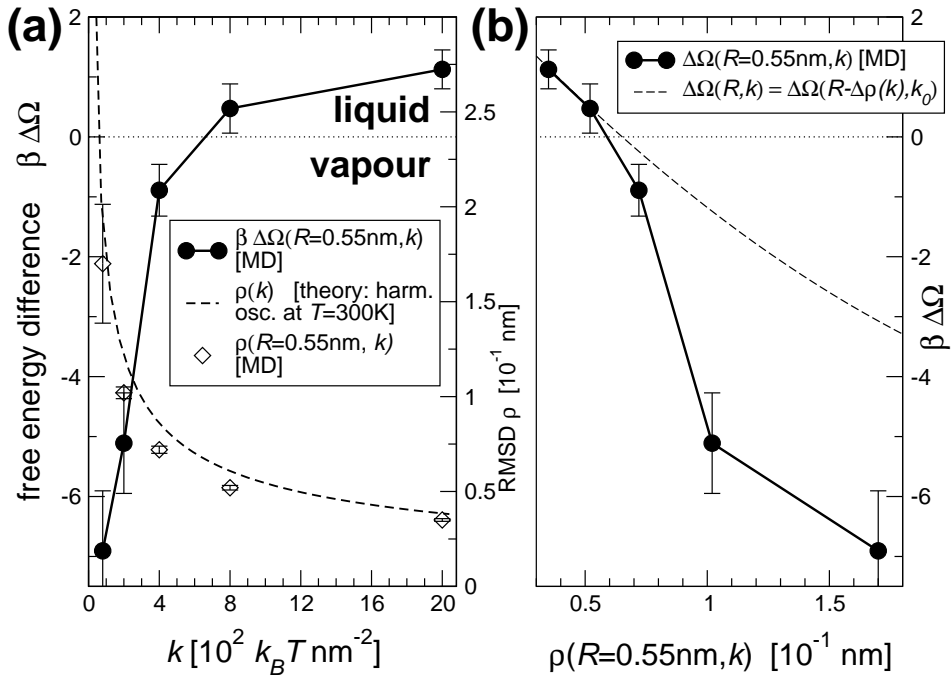}
  \caption{Influence of flexibility on the liquid-vapour equilibrium. (a) For
    a hydrophobic $R=0.55$~nm pore the harmonic restraint force constant $k$
    of the pore wall atoms was varied, resulting in a range of RMSDs $\rho$,
    which show the expected behaviour of thermic harmonic oscillators. The
    free energy difference $\Delta\Omega$ between liquid and vapour states of
    water in the pore shows strong dependence on the local flexibility $\rho$;
    rigid pores contain liquid water whereas flexible ones favour vapour. The
    effect can not be explained by a reduction of the effective pore radius by
    $\rho$ alone [broken line in (b)]. All data at $T=300$~K.}
  \label{fig:squidgy-omega}
\end{figure*}

The influence of local fluctuations in protein structure can be modelled by
changing the harmonic spring constant $k$ that holds the pore wall atoms at
their equilibrium positions. In a simplified picture of a harmonic oscillator
with an average energy $k_{B} T/2$ per degree of freedom the RMSD $\rho$ is
directly related to $k$ by $ \rho(k) = \sqrt{\expect{\Delta
    r_{\mathrm{max}}}^{2}} = \sqrt{{3 k_{B} T}/{k}}$.  This simple model
overestimates the measured RMSDs by only 10--20\% (table~\ref{tab:squidgy}),
showing that the wall atoms behave like almost independent, thermally driven,
harmonic oscillators (figure~\ref{fig:squidgy-omega}a).
%
%
\begin{table*}[tb]
  \centering
  \caption{RMSD of wall atoms in flexible pore models of radius $R=0.55$~nm
    and influence on the liquid-vapour equilibrium of water in the pore at
    $T=300$~K.} 
  \label{tab:squidgy}
  \begin{tabular}{cr@{$\pm$}lcr@{$\pm$}lr@{$\pm$}l}
     \toprule
     restraint force constant $k$  
     & \multicolumn{3}{c}{RMSD $\rho(k)$ [$10^{-1}$~nm]} 
     & \multicolumn{2}{c}{$\expect{\omega}$} 
     & \multicolumn{2}{c}{$\beta\,\Delta\Omega$}\\{}%
     [$10^{2}\times k_{B}T\,$nm$^{-2}$] 
     & \multicolumn{2}{c}{MD} & theory
     & \multicolumn{2}{c}{MD} & \multicolumn{2}{c}{MD}\\
     \midrule
     $0.8$   & $1.7$&$0.3$  & $1.92$ & $0.001$&$0.001$ & $-6.9$&$1.0$ \\
     $2$     & $1.0$&$0.0$  & $1.22$ & $0.006$&$0.005$ & $-5.1$&$0.8$ \\
     $4$     & $0.72$&$0.02$  & $0.86$ & $0.291$&$0.089$ & $-0.89$&$0.43$ \\
     $8$     & $0.52$&$0.01$  & $0.61$ & $0.616$&$0.097$ & $0.47$&$0.41$ \\
     $20$    & $0.35$&$0.01$  & $0.39$ & $0.755$&$0.060$ & $1.1$&$0.3$\\
     \bottomrule
   \end{tabular}
\end{table*}
At a fixed radius, increasing the flexibility (smaller $k$, hence greater
$\rho$) shifts the equilibrium towards vapour (figure~\ref{fig:squidgy-omega}).
Conversely, a more rigid wall favours the condensation of water in the pore.
Thus, water does not ``push away'' the pore walls to fill the pore but rather
fluctuating pore walls appear to disfavour the formation of adjacent water
layers.
This effect cannot be explained with the assumption that the water molecules
encounter a more narrow pore ``on average''. Even if one assumes that the pore
is narrowed by the RMSD of an atom down to an effective radius $R(k) = R_{0} -
\rho(k)$, the effect is still much stronger. When the thermodynamic model and
the parameters for the hydrophobic pore (from table~\ref{tab:tdynmodel} where
$k_{0}=4\times 10^{2} k_{B}T\ \mathrm{nm}^{-2}$) are used to predict
$\Delta\Omega\big(R(k)\big)$ then the prediction compared to the simulation
results underestimates the effect for the more flexible pores but is in
agreement for more rigid pores (figure~\ref{fig:squidgy-omega}b).
We hypothesise that the random positional fluctuations of the wall atoms
change the effective potential between the wall atom and water molecules by
``smearing out'' the attractive well of the interaction potential. Preliminary
calculations showed that an increase in flexibility (or temperature) would always
decrease the well depth of the Lennard-Jones potential, which is used to
describe the interaction between a water molecule and the methane molecules of
the pore wall. \citet{Nijmeijer91} demonstrated the dependence of the surface
tension on the well depth $\epsilon$ of the interaction potential
analytically. MD simulations of a water droplet with 1000 molecules ($R\approx
2.5$~nm) on a graphite-like surface with varying well depth
$\epsilon_{\mathrm{C-OW}}$ yielded a contact angle that varied from
$30^{\circ}$ to $180^{\circ}$ \cite{Lundgren02}.
Thus it is the well depth of the interaction potential which ultimately
determines the thermodynamic equilibrium in the pore. This is made explicit by
considering equation~\ref{eq:DOmega} and noting that $\Delta\Omega(R)$ depends
on the water-wall surface tensions through $\gamma_{vw}-\gamma_{lw}$, which in
turn depend on $\epsilon$. For the case of water in confined geometries it was
shown that capillary condensation and evaporation phenomena depend sensitively
on $\epsilon$ \cite{Hum01,Giaya02,Beckstein03} (or more precisely, on the
effective well depth, the product of the wall atom density with $\epsilon$
\cite{Beckstein03}).

\subsection{NaCl electrolyte}
\label{sec:toypore-ions}

\begin{figure*}[btp]
  \centering
  \newlength{\wions}
  \setlength{\wions}{0.19\linewidth}

  \begin{minipage}{0.85\linewidth}
    \begin{tabular}{@{}lc@{}c@{}c@{}c}
      $R$   & 0.35 nm & 0.6 nm & 0.65 nm & 0.75 nm\\
      \rotatebox{90}{\textcolor{black}{water}}   
        & \includegraphics[clip,width=\wions]{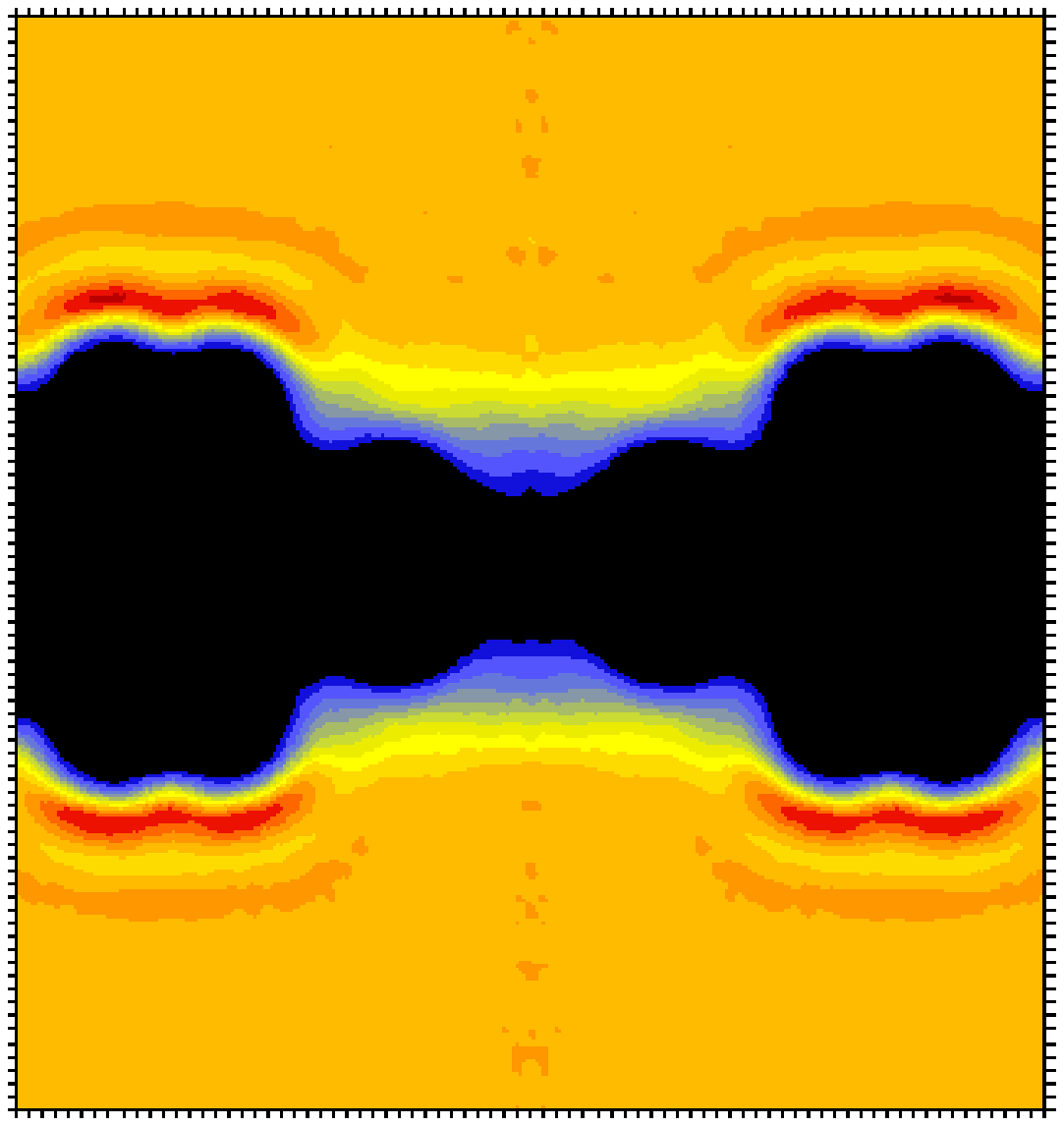} 
        & \includegraphics[clip,width=\wions]{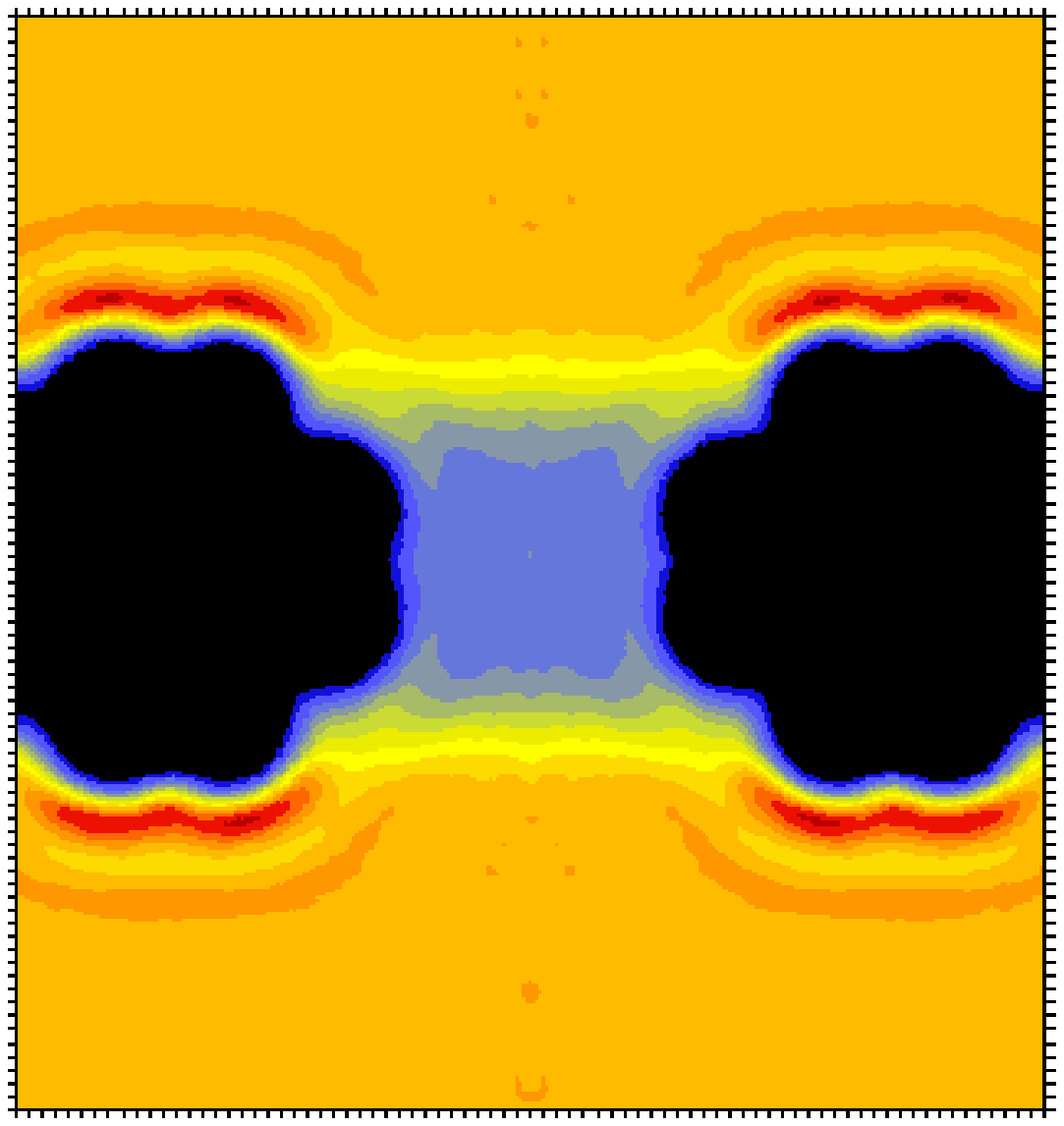} 
        & \includegraphics[clip,width=\wions]{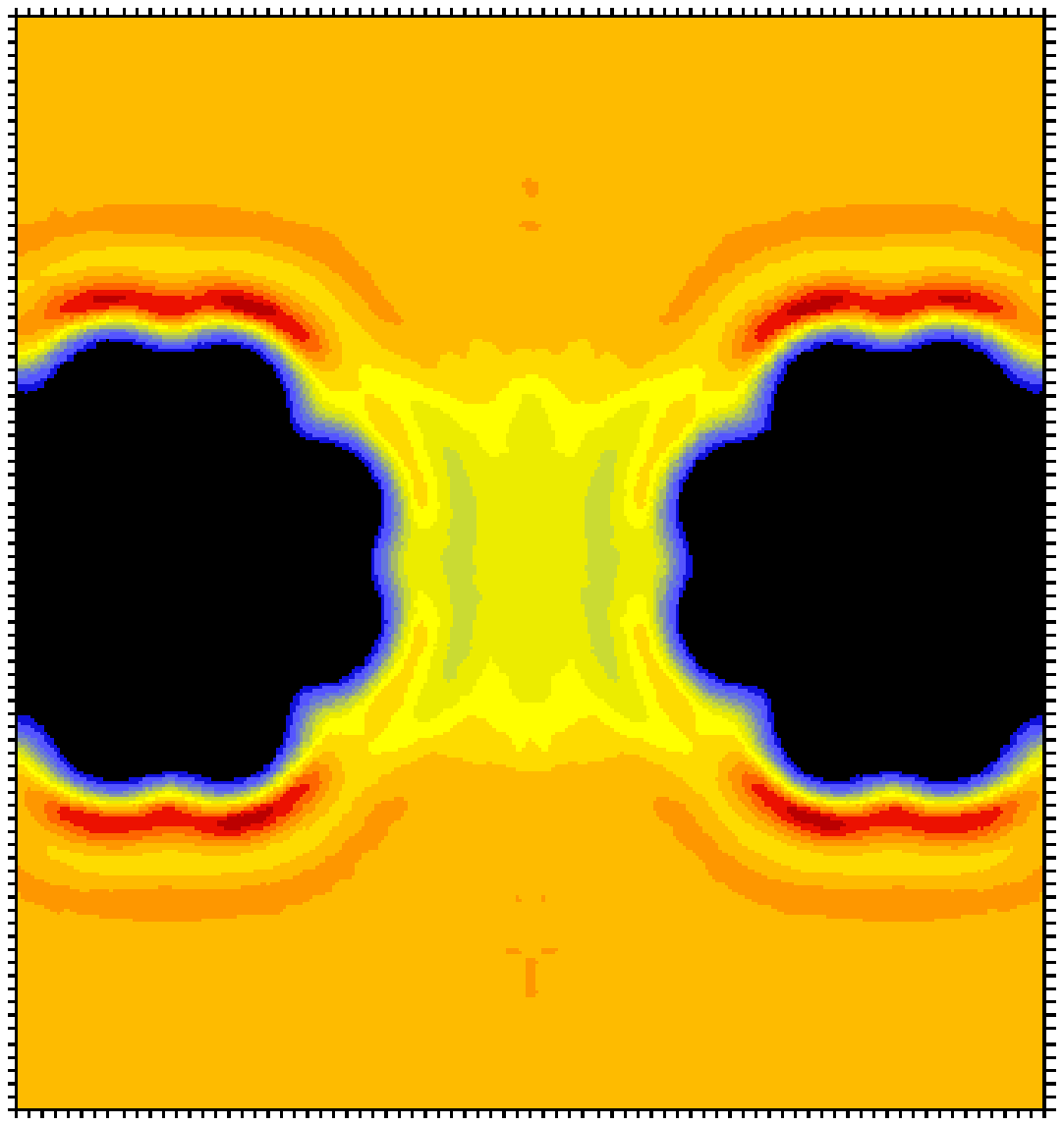}
        & \includegraphics[clip,width=\wions]{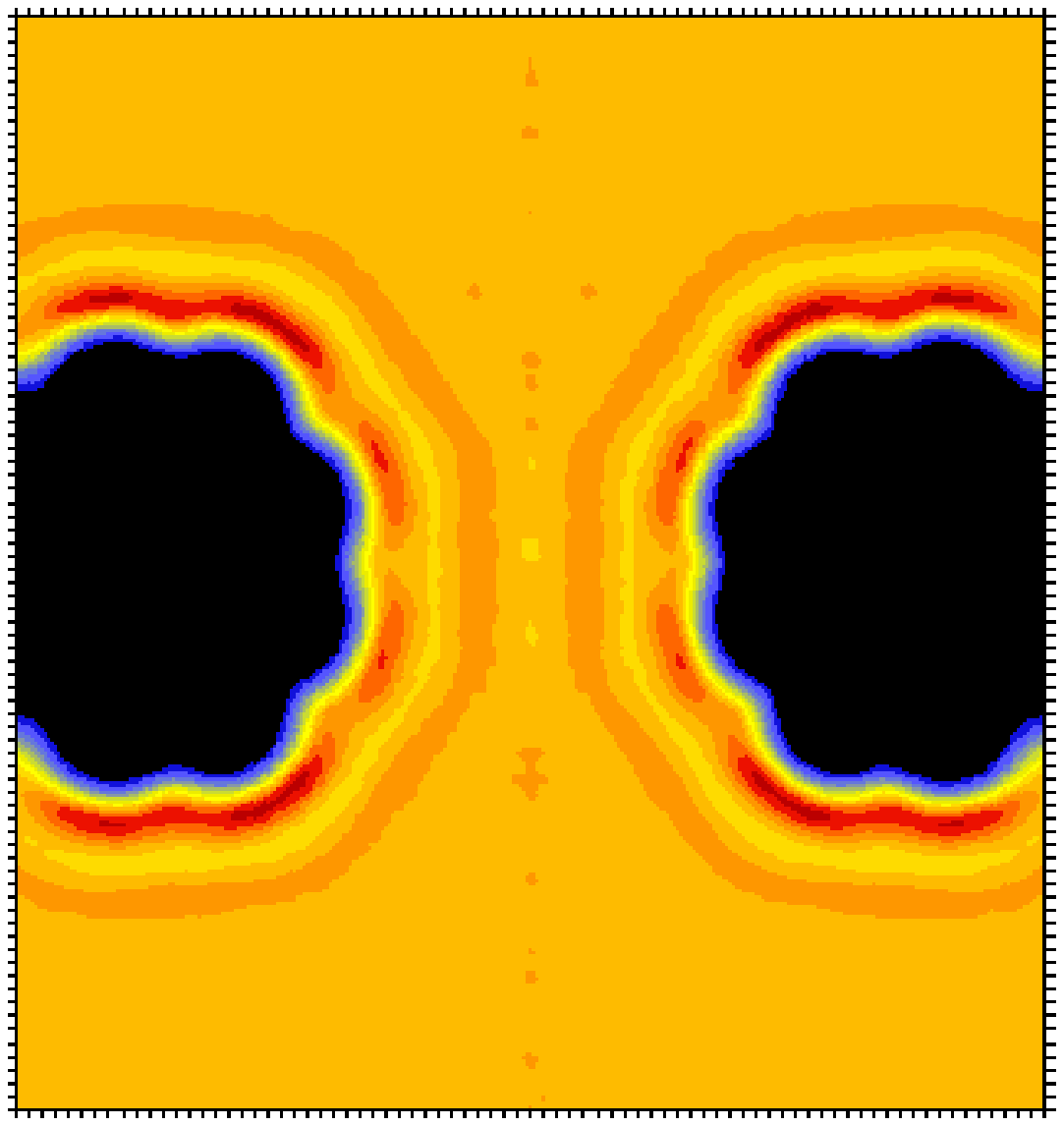}\\
      \rotatebox{90}{\textcolor{black}{$1.3$~M Na$^{+}$}}  
        & \includegraphics[clip,width=\wions]{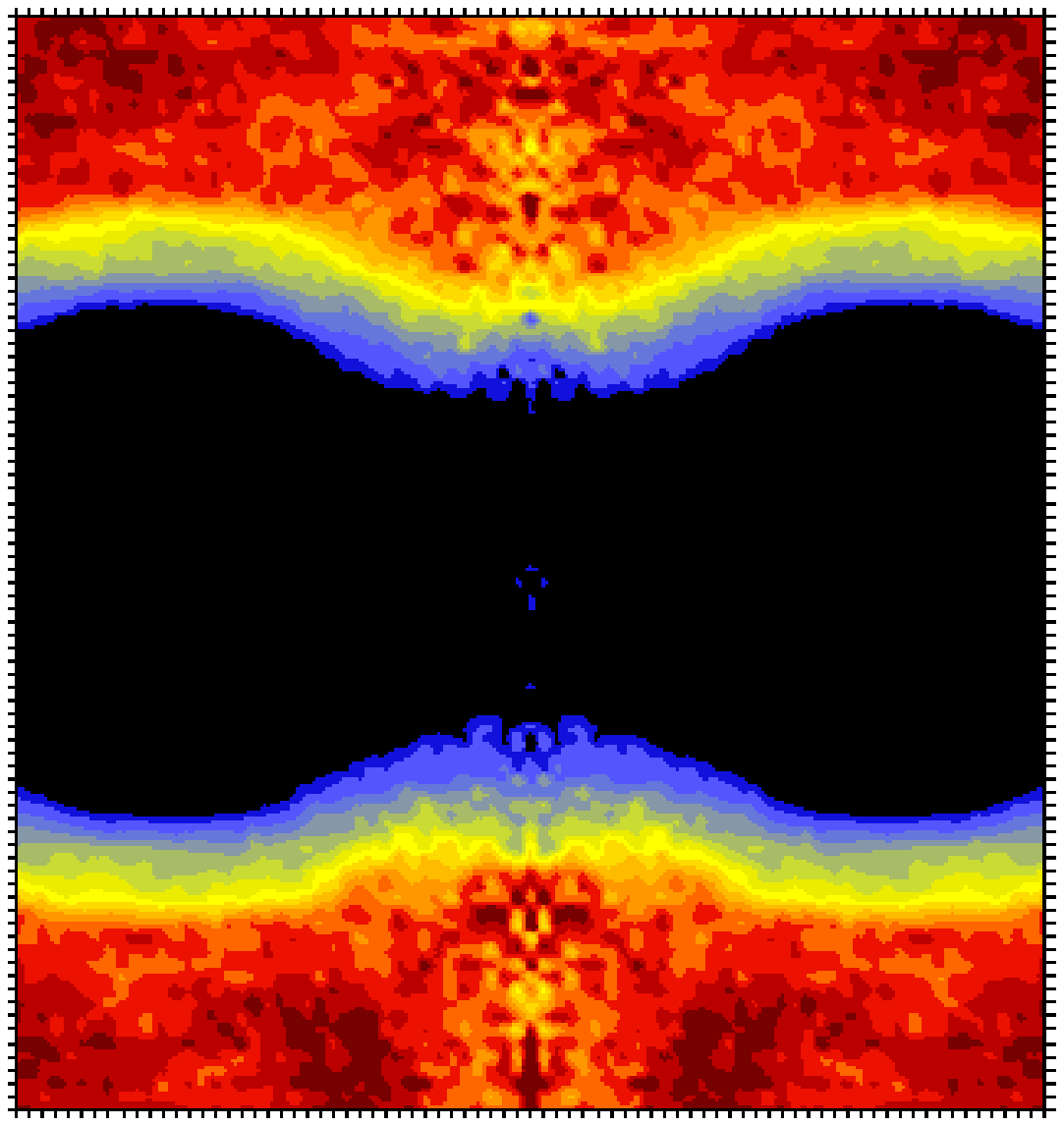} 
        & \includegraphics[clip,width=\wions]{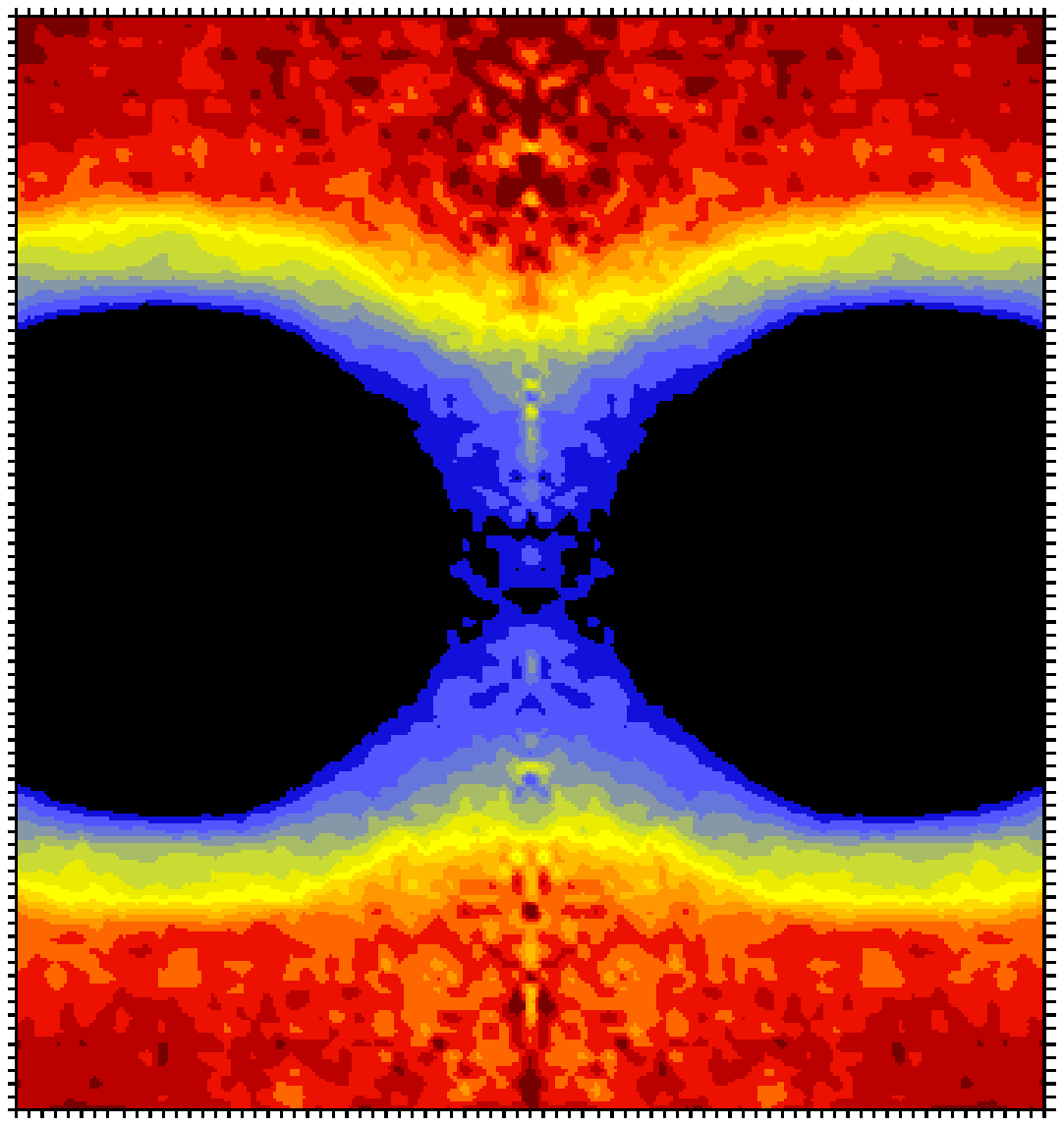} 
        & \includegraphics[clip,width=\wions]{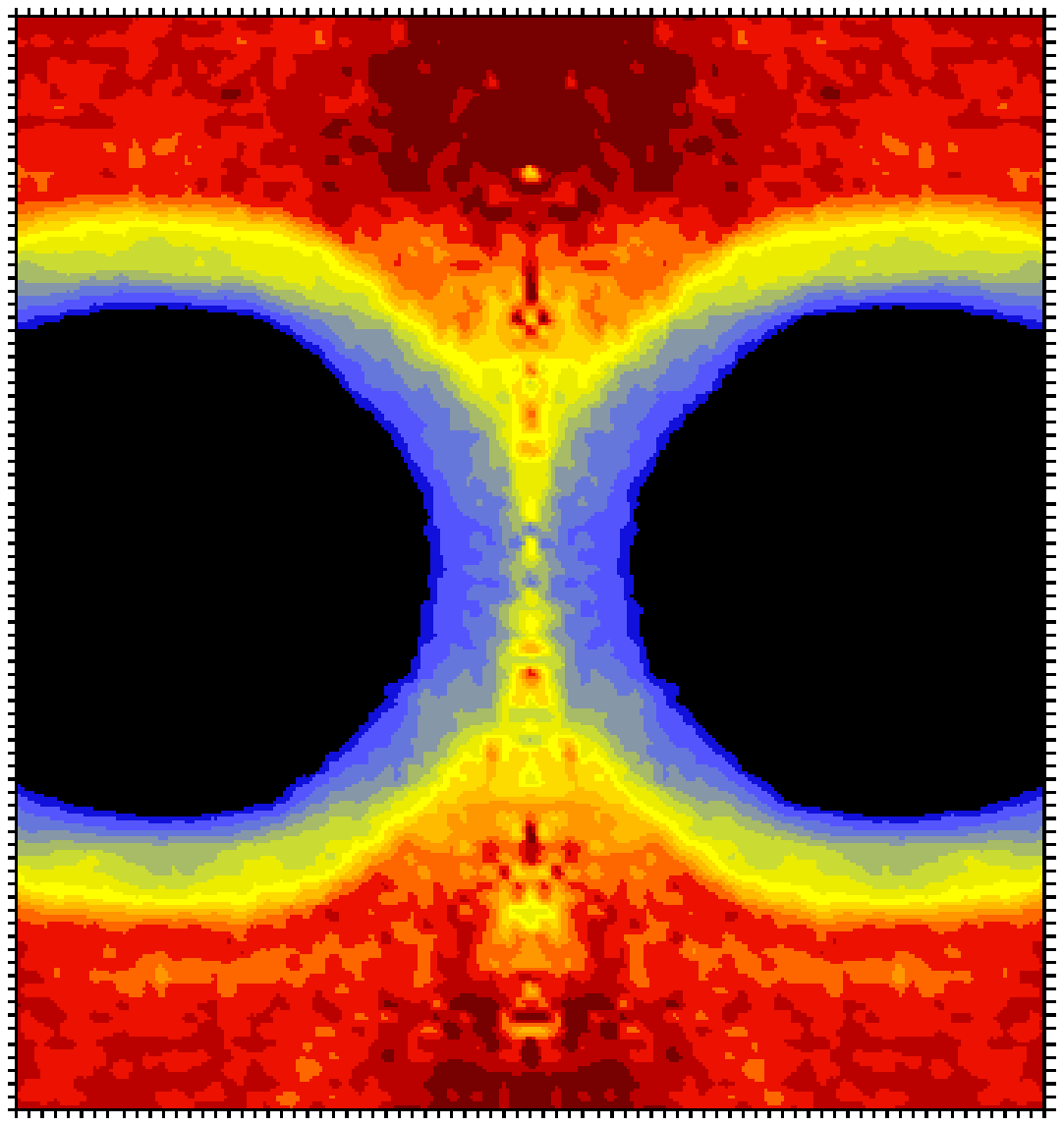} 
        & \includegraphics[clip,width=\wions]{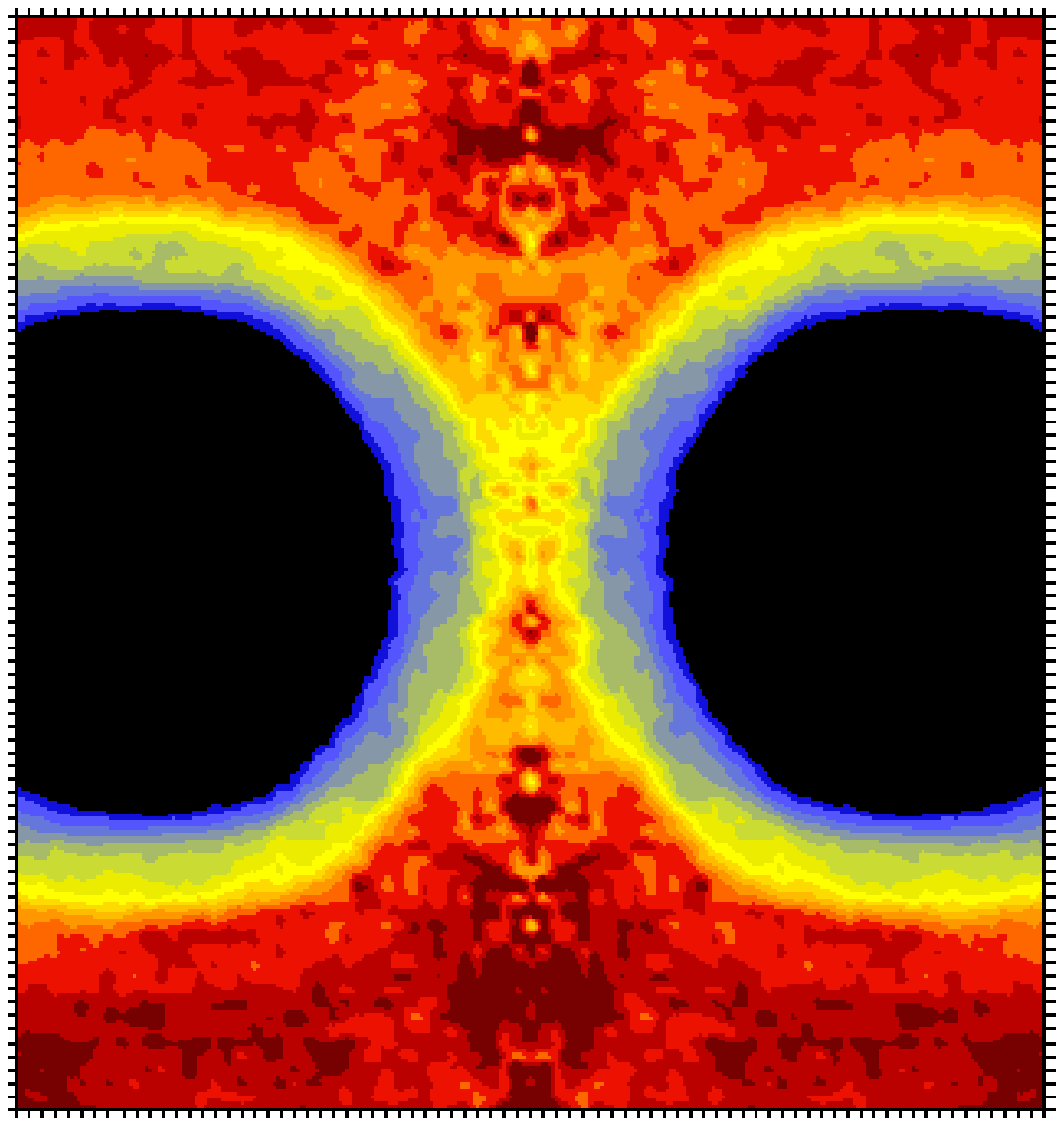}\\
      \rotatebox{90}{\textcolor{black}{$1.3$~M Cl$^{-}$}}
        & \includegraphics[clip,width=\wions]{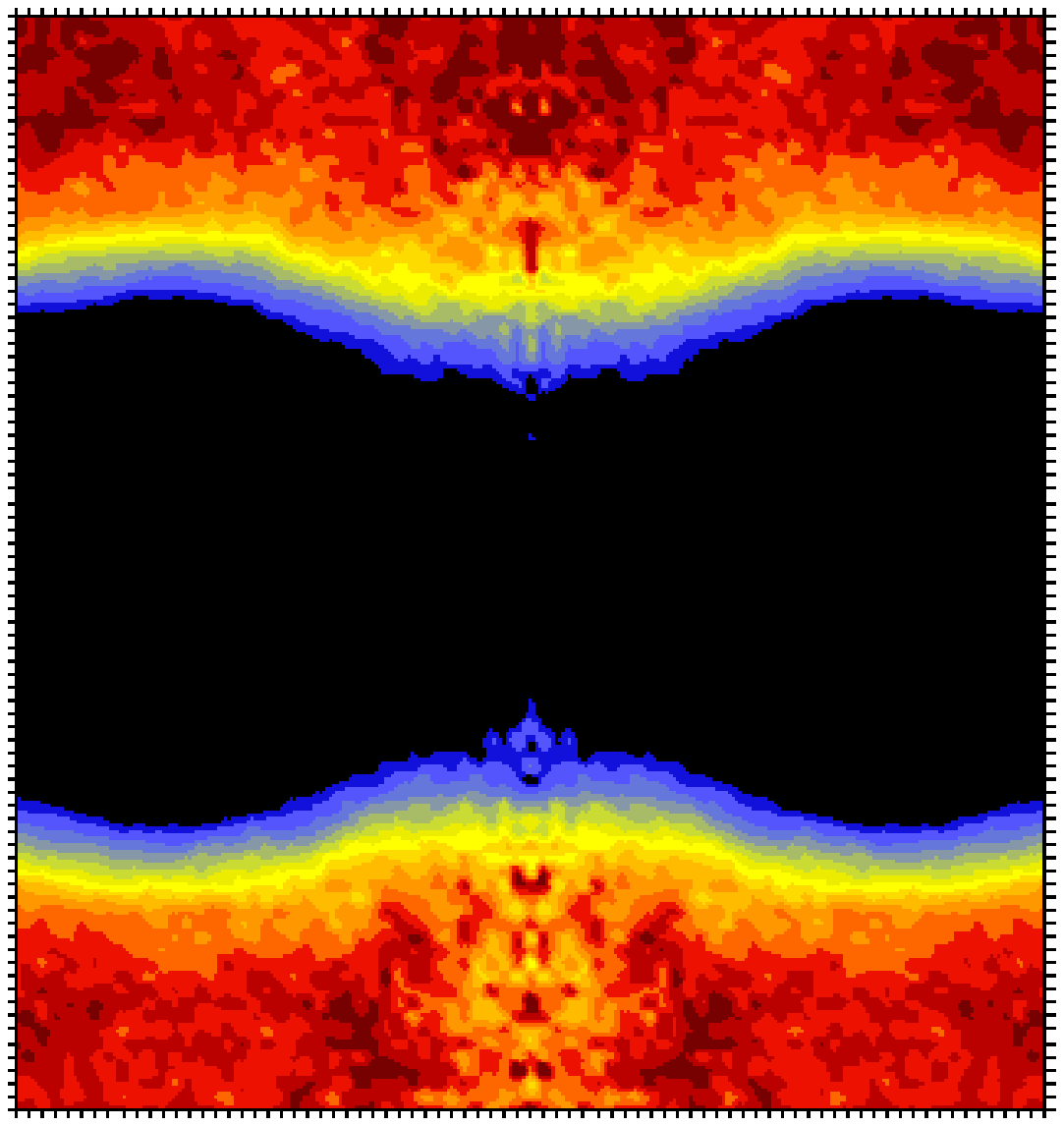} 
        & \includegraphics[clip,width=\wions]{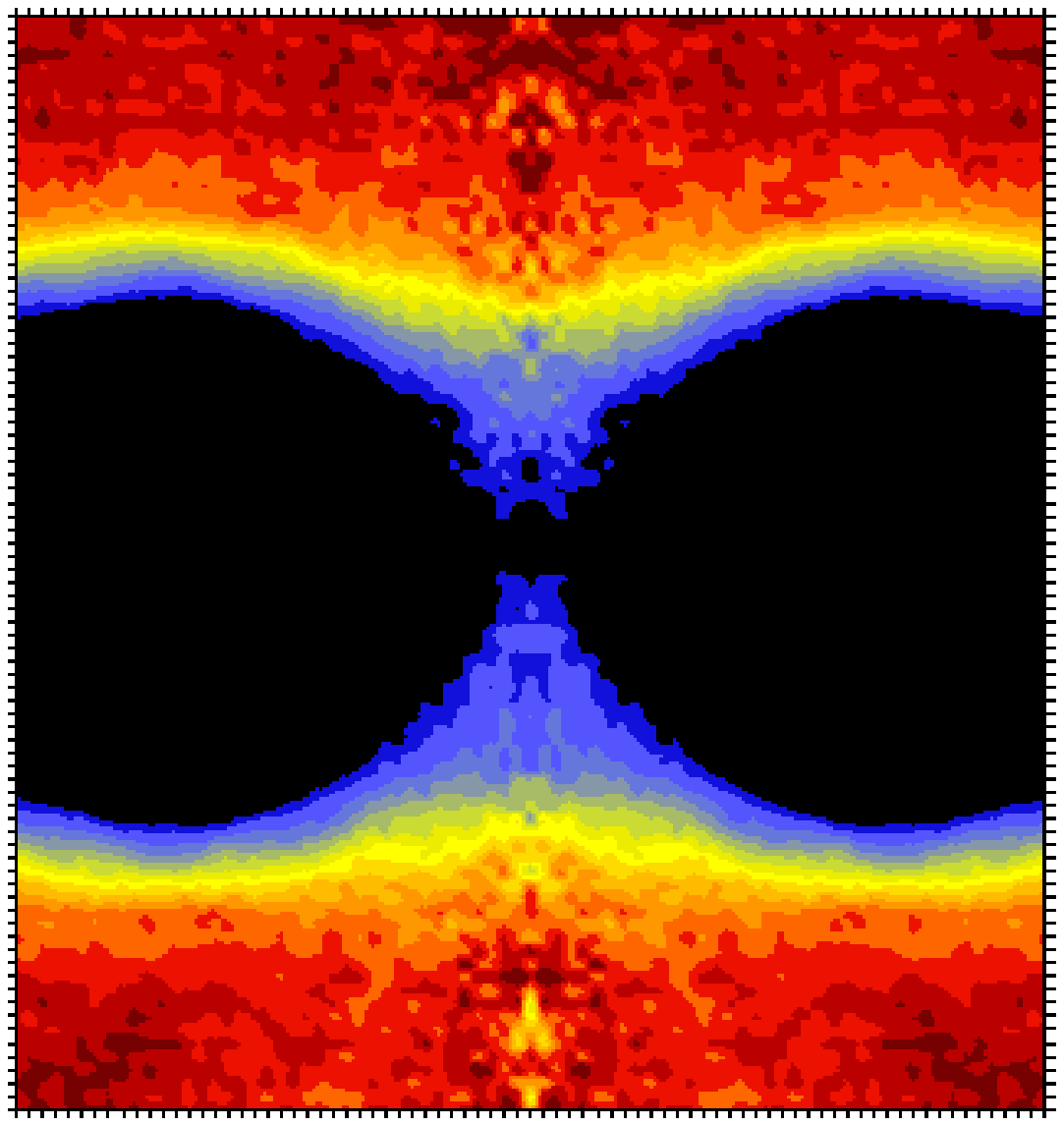} 
        & \includegraphics[clip,width=\wions]{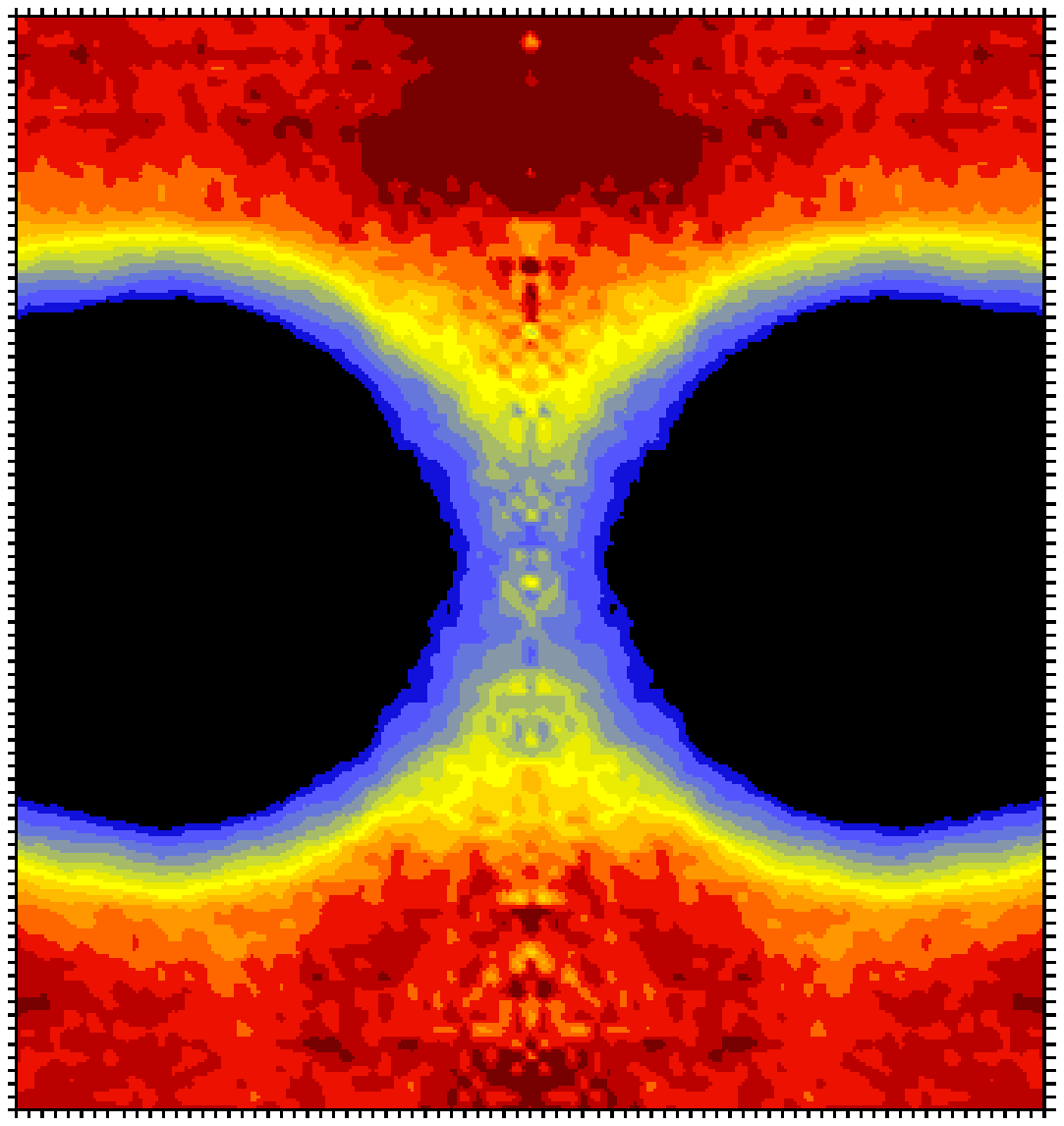} 
        & \includegraphics[clip,width=\wions]{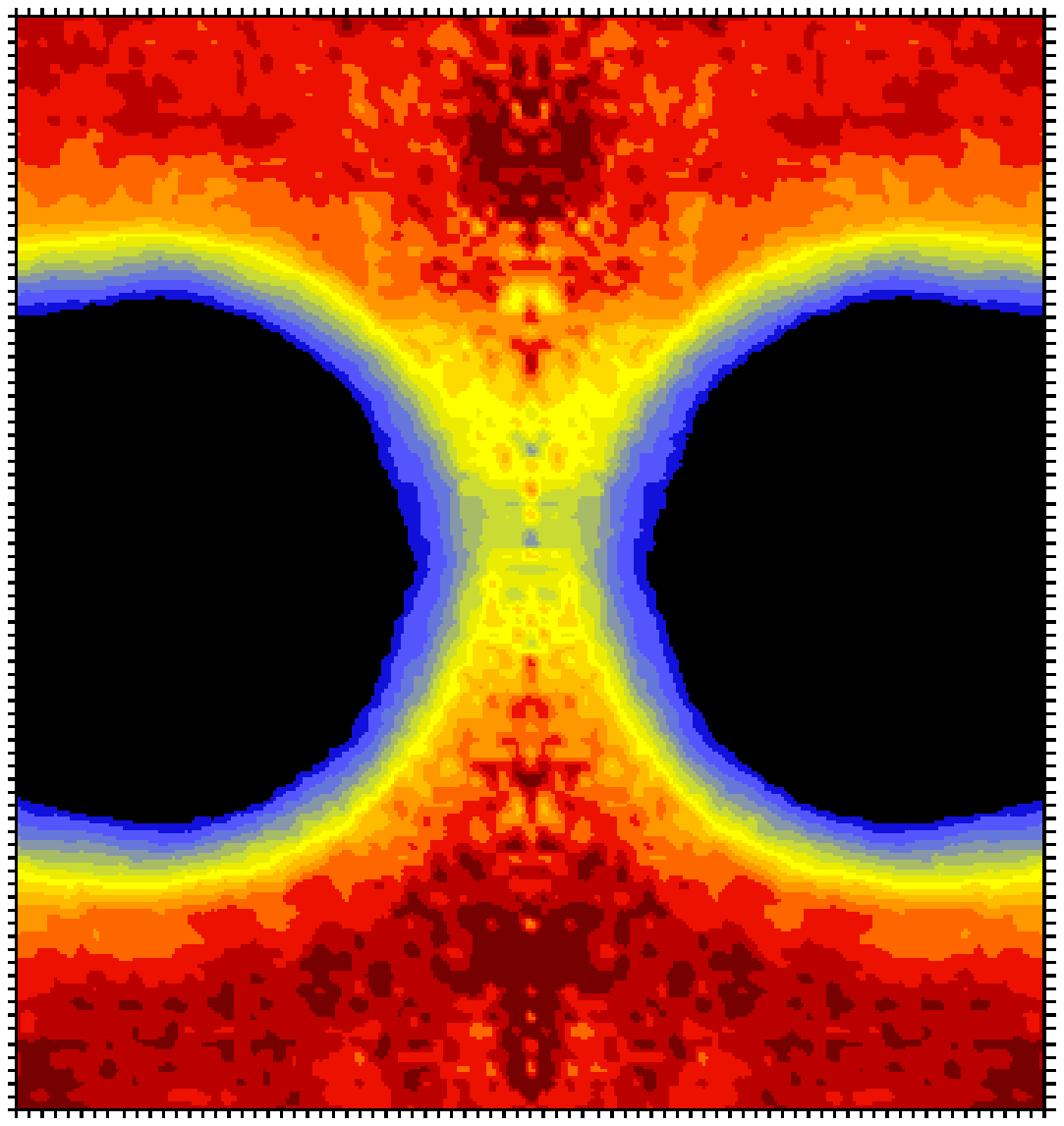} 
     \end{tabular}
  \end{minipage}%
  \begin{minipage}{0.15\linewidth}
    \small density scale:\\
           water: $n_{\mathrm{bulk}}$\\
           ions: mol\,l$^{-1}$\\[1\baselineskip]
    \includegraphics[height=0.4\textheight,keepaspectratio,clip]{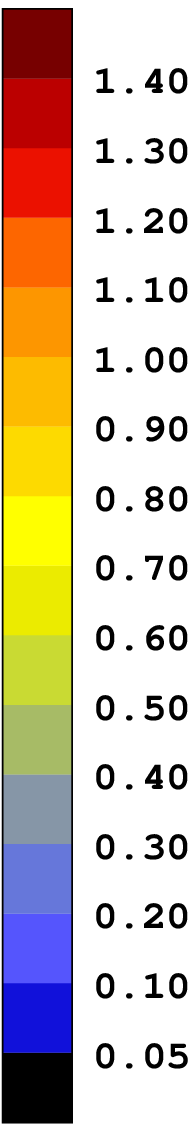}
  \end{minipage}
  \caption[Densities of water, sodium and chloride in model pores]%
  {Equilibrium densities of water (top row), sodium ions (middle) and chloride
    ions (bottom) in hydrophobic model pores of radii $R=0.35$~nm to
    $0.75$~nm; the NaCl bulk concentration is $1.3$~mol\,l$^{-1}$. The narrow
    $0.35$~nm pore mimics the \emph{closed} gate of nAChR whereas the wide
    $0.65$~nm pore approximates the \emph{open} channel. Although wide enough
    to admit the ions physically, the narrow pore is effectively closed to
    ions (``hydrophobic gating''). The membrane mimetic is located in the
    horizontal region which appears black, i.e.{} void of water, in all
    images.}
  \label{fig:ions-rzp}
\end{figure*}

\begin{figure*}[btp]
  \centering
  \includegraphics[clip,width=0.5\linewidth]{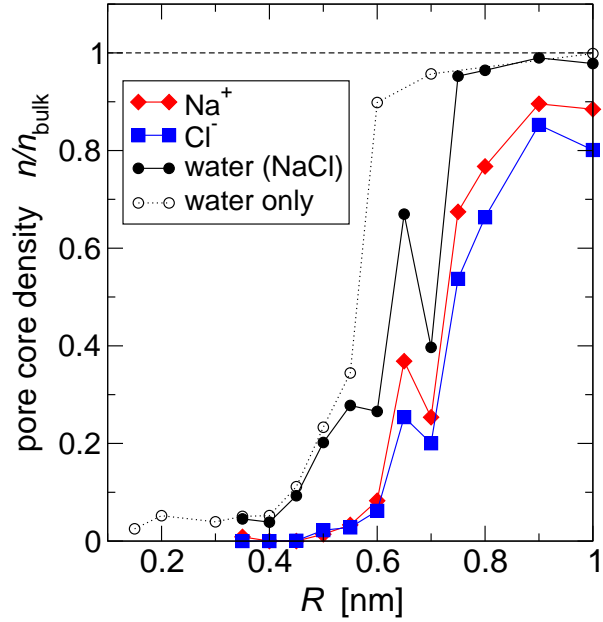}
  \caption{Ionic density in the core region of hydrophobic pores. The core is
      taken to have a radius of about one ionic radius ($0.1$~nm) and exhibits
      the highest density of ions. The core density increases markedly at the
      critical radius $R=0.65$~nm. The core density follows the openness (as
      shown for water in figure~\protect\ref{fig:omegafree}). The dip in density
      at $R=0.7$~nm is due to very few state changes in the $150$~ns simulation
      and a subsequent bias towards a $70$~ns vapour period; in general the
      state switching frequency in the electrolyte systems is reduced compared
      to the pure water case (see also figure~\protect\ref{fig:states}).  Bulk
      concentration of ions is $1.3$~M.  Lines are drawn as guides to the eye.}
  \label{fig:coredensity}
\end{figure*}

As mentioned in the Introduction it has been hypothesised that a local
hydrophobic environment would present a significant desolvation barrier to ion
permeation.  Our simulations of model pores bathed in a $1.3$~M NaCl solution
exhibit a striking change in the behaviour of the ions for pore radii $R \geq
0.65$~nm.  Figure~\ref{fig:ions-rzp} shows that in pores of comparable
dimensions to the closed nAChR pore ($R=0.35$~nm) ions have a vanishing
probability of entering the pore, but at the open-state radius $R=0.65$~nm the
pore density rises to half the bulk value.
%
The average density of ions in the centre of the pore
(figure~\ref{fig:coredensity}) also exhibits a sharp increase near $R=0.65$~nm,
similar to the behaviour of the openness (figure~\ref{fig:omegafree}).
%

The radially averaged densities show two pronounced water layers with an
interlayer distance $d=0.3$~nm near the membrane mimetic.  The ionic density
drops off sharply at the maximum of the outermost water layer and vanishes
completely at the maximum of the water layer closest to the surface.  Sodium
and chloride ions are both present at the same distance from the surface.
Because the membrane mimetic carries no charge and represents a low dielectric
region (as do the less mobile bound water layers) it is energetically more
favourable for the ions to be solvated in the high dielectric region of the
bulk-like water.
The bound water layers are not a consequence of the presence of the ions as
they also form in pure water near a hydrophobic surface with the same
interlayer spacing of $0.3$~nm \cite{Beckstein03}, and in this case even a
third water layer is discernible with $d=0.35$~nm. Water ordering in
cylindrical and slit pores in the absence of ions was also observed in
simulations by other groups
\cite{Lyn96,All99,Mashl03,Allen03b,Spohr98,Striolo03}.
Effectively, ions are excluded from a zone of thickness $0.6$~nm (Na$^{+}$) or
$0.7$~nm (Cl$^{-}$) from the surface. The ion exclusion zone stretches into
the pore without interruption so that at $R=0.65$~nm there is only a narrow
channel of easily displaceable water which can admit ions.  For wider pores,
the ion density in the pore approaches bulk values
(figure~\ref{fig:coredensity}).

\begin{figure*}[tb]
  \centering \includegraphics[width=0.5\linewidth,clip]{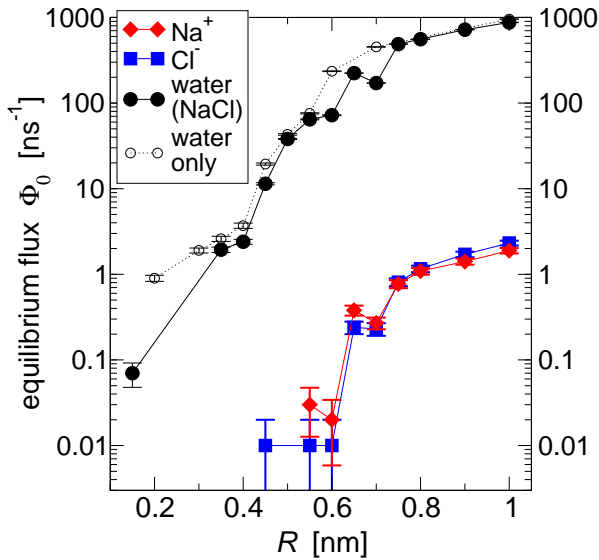}
    \caption{Equilibrium flux $\Phi_{0}$ of water and ions through hydrophobic model
      pores. The total number of ions that successfully permeate the pore per
      nanosecond shows an increase over almost two orders of magnitude when the
      pore radius $R$ is increased from $0.6$~nm to $0.65$~nm. Water flux
      increases with the radius. The most notable influence of the $1.3$~M NaCl
      electrolyte solution is the shift in the transition radius from $0.6$~nm
      (only water) to $0.75$~nm (water with NaCl). Lines are drawn to guide the
      eye and errors are estimated from block averages \protect\cite{Hess99}.}
  \label{fig:ionflux}
\end{figure*}

Although the absence of ionic density can be taken as a strong indication for
a closed pore the presence of equilibrium density does not necessarily imply a
functionally open state, as the latter also requires rapid permeation of ions
through a pore. As MD simulations output the trajectories of water and ions,
the equilibrium flux ($\Phi_{0}$), i.e.\ the total number of particles per
nanosecond which completely permeate the pore, can be measured.
Figure~\ref{fig:ionflux} shows $\Phi_{0}$ for ions and water molecules (both
for water in a $1.3$~M NaCl electrolyte and for pure water) in hydrophobic
pores of varying radius on a logarithmic scale. Water flux is about three
orders of magnitude larger than ion flux and increases rapidly with increasing
radii while $R$ is smaller than the critical radius $R_{c} = 0.62$~nm.  Beyond
$R_{c}$ (when liquid becomes the stable phase) the slope in the logarithmic
plot decreases from $15$~nm$^{-1}$ to $2.4$~nm$^{-1}$. The water current
density $\Phi_{0}/(\pi R^{2})$ reaches a constant value of $300\ 
\mathrm{ns}^{-1}\ \mathrm{nm}^{-2}$ so the increase for $R>R_{c}$ is only due
to the increase in pore diameter.  The increase for $R<R_{c}$ correlates with
the openness (figure~\ref{fig:omegafree}), indicating that the main
contribution to the flux stems from the open-state periods \cite{Beckstein03}.
The presence of ions slightly shifts $R_{c}$ from $0.56\pm0.03$~nm for pure
water to $0.62\pm0.07$~nm for $1.3$~M NaCl electrolyte. Qualitatively, this is
explained by the increase of surface tension with ionic concentration
\cite{MolecDrivingForces03}. This leads to a higher contact angle $\theta_{e}$
\cite{Williams75}, which is also observed in our simulations where it
increases from $134^{\circ}$ to $140^{\circ}$. Hence, $R_{c} = -L
\cos\theta_{e}$ increases for electrolytes compared to pure water.

%
The total number of ions that successfully permeate the pore per nanosecond
shows an increase over almost two orders of magnitude when the pore radius $R$
is increased from $0.6$~nm to $0.65$~nm. Thus the pore ``opens'' at a radius
much larger than the bare ionic radius.
Ions permeate only when the pore is filled with liquid water (typical data in
figure~\ref{fig:states}) and Na$^{+}$ permeation events are not correlated with
Cl$^{-}$ ions passing through the pore.

Because the simulations are in thermodynamic equilibrium the net flux through
the pore is zero for all $R$.  In order to compare the MD equilibrium flux
$\Phi_{0}$, which is determined by the intrinsic free energy barrier to ion
permeation, to experimentally measured non-equilibrium fluxes $\Phi$ of ions
of charge $q$ at a driving transmembrane voltage $V$ we employ rate theory and
estimate
\begin{equation}
  \label{eq:current}
  \Phi(V;R) = \Phi_{0}(R)\, \sinh \frac{q V}{2 k_{B}T}.
\end{equation}
For $R=0.65$~nm, $V=100$~mV, $q=1e$ and $T=300$~K we obtain $\Phi\approx
1.2$~ns$^{-1}$, which is 40 times larger than the experimental value of
$0.03$~ns$^{-1}$ for nAChR at $0.2$~M ionic concentration (calculated from a
conductance of ca.~$45$~pS \cite{Hille01}).  The discrepancy between our
estimate and the experimental data is not unexpected as
equation~\ref{eq:current} has only qualitative character. Furthermore, nAChR
presents a more complex pore lining surface than the model pore, and if the
difference in bath concentrations and access resistance was taken into
account the discrepancy would likely be reduced by an order of magnitude.
Nevertheless, the estimate demonstrates that the intrinsic barrier to ion
permeation is small beyond the critical radius.

\subsection{Sensing external parameters}
\label{sec:sensing}

In the simplest model of sensing a system exists in an equilibrium with two
states.  External stimuli shift the equilibrium which elicits a cellular
response. For example, a temperature-sensing channel at normal temperature
might be in the closed state. Elevated temperature shifts the equilibrium to
the open state and the influx of ions initiates a signalling cascade that
terminates in the sensation of heat. However, the exact mechanisms of gating
by temperature of e.g.\ TRP channels remain obscure \cite{Clapham03}.
%
\begin{figure*}[tb]
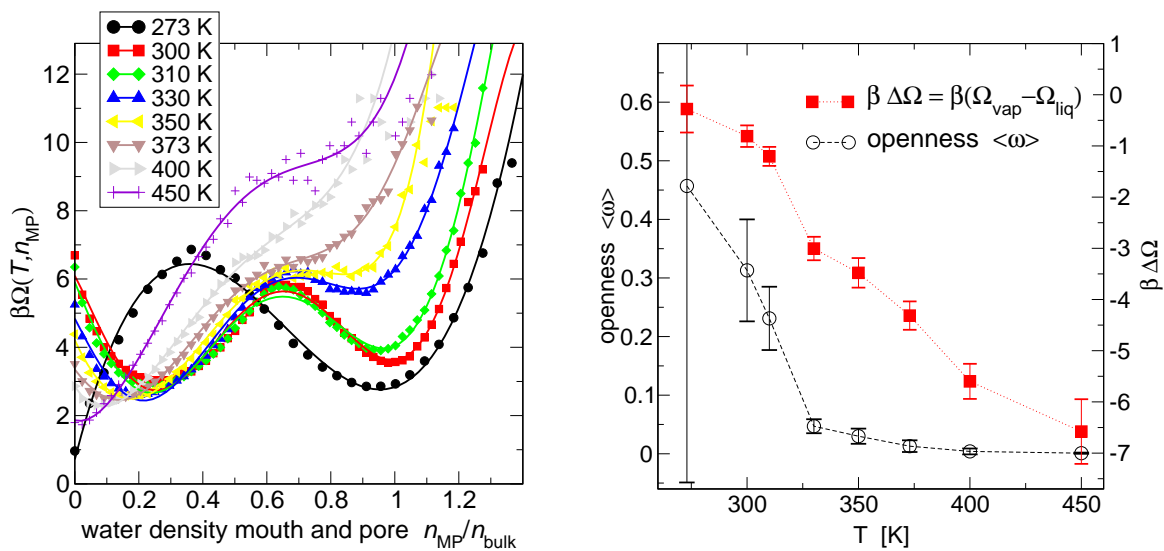

  \centering
  \includegraphics[height=7.2cm,keepaspectratio,clip]{figs/Temperature/rho_free} %
  \hspace{2em}\includegraphics[height=6.9cm,keepaspectratio,clip]{figs/Temperature/Tomega}%
  \caption{Temperature dependence of the liquid-vapour
      equilibrium in a hydrophobic $R=0.55$~nm pore. \emph{Left}: Change of the
      free energy landscape $\Omega(n,T)$ with $T$. The density is calculated
      over the pore and the mouth regions which act as a buffer of constant
      chemical potential except for $T=273~K$, where the mouth region also fills
      with vapour. Continuous lines are polynomial least square fits of order 4
      (or order 6 for $300\ \mathrm{K}\leq T\leq 330\ \mathrm{K}$), reminiscent
      of a Landau free energy with the water density in the pore and mouth
      region $n_{\mathrm{MP}}$ as order parameter. The critical point of SPC
      water is near $T=373$~K. \emph{Right}: Openness and free energy difference
      between states $\Delta\Omega$.}
  \label{fig:temperature}
\end{figure*}
We tested the influence of temperature in the range from $273$~K to $450$~K on
water in the $R=0.55$~nm hydrophobic pore (data points for $T>373$~K represent
``supercritical'' water because the critical point for the SPC water model is
close to $T=373$~K). The free energy landscape of the water-pore system
depends on the temperature (figure~\ref{fig:temperature}a).  With increasing
temperature the liquid state (i.e.\ near water bulk density) decreases in
stability. Accordingly, the openness decreases with temperature so that the
vapour state, i.e.{} the closed stated, dominates at higher temperatures
(figure~\ref{fig:temperature}b).
The decrease in openness with increasing $T$ indicates that the difference of
the wall surface tensions $\Delta\gamma_{w}$ must become even more negative,
i.e.\ vapour in contact with the wall is increasingly favoured. The effect
would be similar to the one discussed for increased flexibility of wall atoms
as the major effect of temperature would be to smear out the interaction
potential between water molecules and wall atoms.
This example demonstrates in principle how an external signal, an increase in
temperature from $T=300$~K to $330$~K, decreases the openness, and
correspondingly the flux of water molecules, by a factor of six.

\section{Conclusions and outlook}
\label{sec:conclusions}

We have explicitly demonstrated a hydrophobic gating mechanism for ions, using
simplified hydrophobic pores as models for the closed gates of ion channels.
There is a critical radius $R_{c}$ above which a pore becomes effectively
permeable to water or ions. $R_{c} \approx 0.56$~nm for water and
$R_{c}\approx 0.65$~nm for ions, is much larger than the radius for a water
molecule or a bare ion alone. This correlates nicely with the current view
that the closed state structures of many ion channels contain gates formed by
hydrophobic constriction sites. In nAChR the radius of the putative gate is
$0.31$~nm, $0.17$~nm in MscL, and $0.13$~nm in KcsA. Models of KcsA and of Kv
channels in their open states, based on the MthK and KvAP structures
respectively, have gate radii of about $0.6$~nm \cite{Beckstein03a}, whereas
the open-state nAChR structure opens up to about $0.65$~nm. Our simulations
show that the polarity of the pore wall can shift the critical radius
considerably.  This is reflected in the proposed gating mechanism of nAChR
\cite{Unw00}. Not only does the radius of the constriction site increase but
hydrophobic sidechains are also rotated out of the pore to expose the more
polar, i.e.\ hydrophilic, peptide backbone.  By combining hydrophobic gating
with a change in surface polarity only a moderate change in radius is required
to obtain a large physiological effect.
Changes in local flexibility may also modulate the gating behaviour.  It is
too early to ascertain their importance as such changes in flexibility have
not been widely investigated experimentally for channels (or any other
proteins). There are suggestions of possible regulatory roles of changes in
flexibility in e.g.\ KTN domains \cite{Roosild02}, and in the modulation of
binding of TRAP protein to the Trp operon mRNA \cite{McElroy02}.
Temperature can also affect gating, in a fashion similar to local flexibility,
but its direct effects seem too small to explain, for instance, the
temperature sensitivity of TRP channels on their own.

A thermodynamic model based on surface energies fits remarkably well the data
of the atomic-scale MD simulation, even for very small radii.  Though such a
macroscopic treatment is not \emph{a priori} expected to give a satisfactory
description of a microscopic system there are other examples as for instance
classical nucleation theory, which shares some similarities with our and
others' models \cite{Allen03b,Huang03}. It generally agrees quite well with
experiments on the condensation of droplets from vapour, which implies that
the use of macroscopic surface tension is valid even for droplets of radii of
about $1$~nm \cite{Adamson90}.  It appears that in this case water structure
(i.e.{} a hydrogen bond network) is only important insofar that it is
responsible for different wall-fluid surface tensions.  Our results seem to
corroborate the conclusions of \citet{Maibaum03} that two-state behaviour of
water in pores should only require a cold liquid close to phase coexistence
and sufficiently different vapour-wall and liquid-wall surface tensions (which
is, of course, where the ``special'' properties of water reside in such a
model).

We have sketched out principles of gating mechanisms in ion channels, based on
model channels. Ion channels in nature display complex conformational dynamics
in relationship to gating. There is a need for better single molecule methods
for ion channels to probe these phenomena experimentally. Combined electrical
and optical methods look promising in this respect \cite{Harms03} although
considerable improvements in time resolution are still required. In the
meantime there is a continued role for simulations and theory to enable us to
bridge between static structure and dynamic function.

In addition to hydrophobic gating \emph{per se}, our studies suggest the
possible importance of gate flexibility in regulation of biological activity.
The relationship between static structures, conformational change, and
intrinsic flexibility in relationship to proteins and signalling merits
further active investigation. In a recent review of protein-protein
interactions and conformational changes, \citet{Goh04} conclude that there is
increasing support for a pre-existing equilibrium model. In such a model,
proteins exist in a population of conformations, with ligand binding leading
to a change in the probability distribution of the ensemble. Some evidence in
support of this have been obtained from recent simulation studies of ligand
binding proteins (e.g. \cite{Pang03}) but further (single-molecule)
experimental and computational studies of a wider range of proteins are needed
to more firmly establish the general importance of changes in flexibility.

\paragraph{Acknowledgements}
We are grateful for the interest and encouragement of all our colleagues and
for discussions with Paul Barrett, Kaihsu Tai, Campbell Millar, Jos{\'e}
Faraldo-G{\'o}mez, Beno{\^{\i}}t Roux and Nigel Unwin.  This work was funded
by The Wellcome Trust.

\widetextNR 

\begin{multicols}{2}
[\paragraph{Abbreviations list}]
\begin{description}
  \small \addtolength{\itemsep}{-0.5\baselineskip}
\item[MD] molecular dynamics
\item[SPC] simple point charge water model
\item[RMSD] root mean square deviation
\item[nAChR] nicotinic acetylcholine receptor
\item[MscS] mechanosensitive channel of small conductance (\emph{Escherichia
    coli})
\item[MscL] mechanosensitive channel of large conductance (\emph{Mycobacterium
    tuberculosis})
\item[KcsA] potassium channel (\emph{Streptomyces lividans})
\item[KirBac1.1] inward rectifier potassium channel (\emph{Burkholderia
    pseudomallei})
\item[MthK] calcium-gated potassium channel (\emph{Methanobacterium
    thermoautrophicum})
\item[KvAP] voltage-dependent potassium channel (\emph{Aeropyrum pernix})
\item[D2] model pore containing two dipoles (amphipathic)
\item[D4] model pore containing four dipoles (hydrophilic)
\item[TRP] transient receptor potential channel family
\item[KTN] potassium transport, nucleotide binding domain
\item[TRAP] trp RNA-binding attenuation protein
\item[trp] tryptophan
\item[mRNA] messenger RNA

\end{description}

\section*{Glossary}
\begin{description}
\small
\item[Classical molecular dynamics] A computational scheme to calculate the
  time evolution of a microscopic system that is treated in atomic detail.
  Interactions are parametrised through classical two, three or four body
  potentials. A collection of these parameters is known as a ``force field''.
  The dynamics are obtained by integrating Newton's equations of motion.
\item[Gating] An ion channel can exist in an ``open'' state, which allows for
  ion permeation, and a ``closed'' state, when ions cannot pass the channel's
  gate and thus are prevented from crossing the membrane.  Such a channel can
  be switched between states by external signals.
\item[Surface tension, surface free energy] The free energy per area required
  to create an interface between two phases. It is always positive, i.e.{} it
  always costs free energy to create a surface compared to the homogeneous
  bulk phases.
\item[Contact angle] The angle which is formed between the tangent on a
  droplet of liquid where it is contact with the surface and the surface
  itself. The droplet is taken to be in equilibrium with its vapour. A surface
  material is called hydrophobic (``water-hating'') when the contact angle
  with water is greater than $90^{\circ}$ (the drop sits on the surface) and
  hydrophilic (``water-loving'') when it is smaller (the drop resembles a
  pancake).
\end{description}

\end{multicols}

%
%
  \small 

\end{document}